\documentclass[12pt,journal, onecolumn, draftcls]{IEEEtran}

\usepackage{amsmath,graphicx}
\usepackage{float,bm}
\usepackage{verbatim}
\usepackage{mathrsfs}
\usepackage{cite}
\usepackage{amsthm}
\usepackage{epstopdf}
\usepackage{epsfig,psfrag}
\usepackage{amsfonts,amssymb}
\usepackage{color}

\newtheorem{theorem}{Theorem}

\newtheorem{proposition}{Proposition}

\def\var{{\mbox{var}}}
\def\cov{{\mbox{cov}}}
\def\ave{{\mbox{ave}}}
\def\E{{\mathbb{E}}}
\def\diag{{\mbox{diag}}}

\newtheorem{definition}{Definition}

\usepackage{verbatim}
\usepackage{mathrsfs}
\usepackage{float}
\usepackage{tikz}

\begin{document}
	
	\title{Modelling Graph Errors: Towards Robust Graph Signal Processing} 
	
	\author{Jari Miettinen,~\IEEEmembership{Member,~IEEE}, Sergiy~A.~Vorobyov,~\IEEEmembership{Fellow,~IEEE}, and Esa Ollila,~\IEEEmembership{Senior Member,~IEEE}
		\thanks{The authors are with the Dept. Signal Processing and Acoustics, Aalto University, PO Box 13000 FI-00076 Aalto, Finland, Emails: {\tt jari.p.miettinen@aalto.fi, sergiy.vorobyov@aalto.fi, esa.ollila@aalto.fi}. Some preliminary results of this work have been presented at the {\it IEEE ICASSP 2018}. This work was supported in parts by the Academy of Finland grants No.~299243 and No.~298118. Sergiy~A.~Vorobyov is the corresponding author.}
	}
	
	\maketitle

\begin{abstract}
The first step for any graph signal processing (GSP) procedure is to learn the graph signal representation, i.e., to capture the dependence structure of the data into an adjacency matrix. Indeed, the adjacency matrix is typically not known a priori and has to be learned. However, it is learned with errors. A little attention has been paid to modelling such errors in the adjacency matrix, and studying their effects on GSP methods. However, modelling errors in the adjacency matrix will enable both to study the graph error effects in GSP and to develop robust GSP algorithms. In this paper, we therefore introduce practically justifiable graph error models. We also study, both analytically when possible and numerically, the graph error effect on the performance of GSP methods in different types of problems such as filtering of graph signals and independent component analysis of graph signals (graph decorrelation).  
\end{abstract}
%

\section{INTRODUCTION}
\label{sec:intro}

In the classical signal processing setup where the digital signals are represented in terms of time series or vectors of spatial measurements (for example, measurements by sensor arrays), it is assumed that each point of a discrete signal depends on the preceding or spatially close point of the signal. With the current advances in the data collection and representation, when the data points may not be ordered temporally or spatially and the digital signal may no longer be accurately or even adequately represented by a simple time series structure, the classical signal processing tools are no longer applicable. Thus, graph signal processing (GSP) \cite{SandryhailaMoura2013, Shumanetal2013, Ortegaetal2017} has emerged as a new area of signal processing and data analysis where one of the major aims is to generalize the standard signal processing methods and concepts into the context of more complex signals represented on graphs that can have various structures. The graph corresponding to ordered data is then just a simple special case of a digital signal graph. Examples of digital signals represented on graphs include sensor networks, brain networks, gene regulatory networks, and social networks to name just a few most frequently encountered graph signals \cite{TuSayed2011, Acemogluetal2013, Huangetal2016, Blochletal2010}.

In GSP, each signal value is indexed by a node in a graph at which this signal value is measured, and edges between pairs of nodes in a graph indicate dependencies between the corresponding signal values. Then the underlying graph can be fully specified by an adjacency matrix, denoted hereafter as $\mathbf{A}$, whose $(i,j)$th element is nonzero if the $i$th and the $j$th nodes are connected, and the value $[\mathbf{A}]_{i,j}=a_{ij}$ describes the strength of the relationship. The adjacency matrix can be viewed as the basis of GSP, on which for example the graph Fourier transform (GFT) and graph filters \cite{SandryhailaMoura2013} are built, often via the graph Laplacian matrix or other {\it shift operators} which are based on the adjacency matrix.\footnote{The GSP literature is developing fast and by now covers also sampling theory for graph signals~\cite{Chenetal2015, Marquesetal2016a}, stationarity theory for graph signals~\cite{Marquesetal2017}, and percolation~\cite{Segarraetal2016} of graph signals. Graph filters have been applied for example in distributed average consensus problem~\cite{SandryhailaKarMoura2014,SafaviKhan2015}.}

The process of obtaining the adjacency matrix may take different forms. There may exist physical or friendship connections between the nodes known in advance, which directly yield a graph. At the other extreme end, there may be no auxiliary information available about the connectedness of the nodes, and the methods for finding an adjacency matrix are then entirely based on training data. Methods for estimating the graph Laplacian matrix in the latter case have been developed in \cite{Dongetal2016, Chepurietal2017, Traganitisetal2017,Pasdeloupetal2018,Segarraetal2017}. Third, the choice of the adjacency matrix may be based on other variables, which are not perfectly correlated with the graph signal which we are interested in. In both the second and the third cases, the process of adjacency matrix learning is necessarily stochastic. As a result, there are multiple sources of errors which lead to imperfect learning of the adjacency matrix. Even in the case when the choice of the adjacency matrix seems to be obvious and based on the known physical or friendship connections between the nodes, such choice still may not be the best one. 

Our focus in this paper is to study consequences of imperfect specifications of graph signal adjacency matrix to GSP tasks. The source of errors in the adjacency matrix learning may be related to or be completely irrelevant to the sources of errors in the signal measurements. However, even when the sources of errors in the adjacency matrix learning and signal value measurements are the same, the effect of errors in the adjacency matrix on GSP tasks are different from those caused by the errors in signal measurements. Considerations of the graph learning errors has only recently gained some attention. For example, a discussion on the topic is given in~\cite{CeciBarbarossa2020}, where eigen-decomposition of the graph Laplacian was analyzed when small subsets of edges are added or deleted. Based on the analysis, robust methods were built for GSP tasks such as signal recovery, label propagation and clustering, when the knowledge of edgewise probabilities of errors in the graph topology is available.

There are also some works which study robustness with respect to graph topology errors in specific tasks. For example, total least-squares approach for robust graph signal recovery was proposed in~\cite{Cecietal2020}. Graph neural networks were found to be stable to small changes in graph topology in \cite{GamaBrunaRibeiro2019}. Node-variant graph filters were introduced in~\cite{SegarraMarquesRibeiro2017}, and they were shown by numerical experiments to be more robust to errors in graph shift operator than node-invariant filters. For distributed average consensus problem, \cite{KruzickMoura2017} proposed a method which optimizes worst case performance when the eigenvalue distribution is considered uncertain. In~\cite{SegarraRibeiro2016}, a new centrality measure was introduced and the robustness with respect to edge weight perturbation was considered. The robustness to some missing edges was also part of the comparison of autoregressive moving average graph filters and finite impulse response graph filters for time-varying graph signals in~\cite{Isufietal2017b}. For spectral graph filters, \cite{LevieIsufiKutyniok2019} defined Caley smoothness filter space where the filter perturbation is bounded by a constant times the perturbation of the graph. In the presence of erroneous edges, \cite{CaiLiangRakhlin2017} studied differences between small world graph and Erd\"{o}s--R\'{e}nyi graph, and \cite{Abbeetal2014} studied clustering of nodes into two communities.   

The starting point for studying GSP performance under the condition of imperfect knowledge of graph signal adjacency matrix is the modelling of graph errors. As the errors in the signal modelling in the classical signal processing may lead to significant performance degradation (see  for example \cite{Me2003}), the errors in the adjacency matrix may have a significant effect on the GSP performance. Then the development of robust GSP methods, just as the development of robust signal processing methods \cite{Me2003}, is of a great importance. This paper, to the best of our knowledge, is the first attempt towards developing justifiable and generic enough models for adjacency matrix mismatches. It also studies the effects of the mismatched adjacency matrix on the performance of some more traditional GSP applications such as filtering of graph signals as well as independent component analysis (ICA) of graph signals (also referred to as graph decorrelation (GraDe) in the context of separating signals based on graph structure only). 

\subsection{Contributions}
Our main contributions are the following.\footnote{Our preliminary work on the topic has been reported in \cite{JariICASSP2018}.}
\begin{itemize}
\item We formulate graph error models for the adjacency matrix, which help to quantify the deviation from the true matrix using a few parameters. We consider adding and removing edges and perturbing the edge weights. First, we build models assuming equal probabilities of mislearning the edge status for each pair of nodes, but later we generalize to the case where there are several subsets of pairs such that within the groups the probabilities are equal, but they differ between the groups. 

\item We study the structural effects of the proposed error models on the adjacency matrix, including spectrum analysis where applicable and possible.

\item We introduce the graph autocorrelation, which is a scale and edge density invariant measure of smoothness between the signal and the graph. 

\item We illustrate what effects different type of errors in adjacency matrix specification produce in filtering of graph signal and ICA of graph signals. Some theoretical studies as well as numerical and real data-based studies are performed.   
\end{itemize}
   
\subsection{Notation} 
We use boldface capital letters for matrices, boldface lowercase letters for vectors, and capital calligraphic letters for sets. The exceptions are $\mathbf{1}_N$ which is the $N$-dimensional vector full of ones, the $M \times N$ matrix full of ones $\mathbf{1}_{M\times N} = \mathbf{1}_M\mathbf{1}_N^\top$, and  $\mathbf{1}_\mathbf{A}$ is a matrix of the same size as $\mathbf{A}$, such that $[\mathbf{1}_\mathbf{A}]_{i,j} = 1$, if $a_{i,j}\neq 0$ and $[\mathbf{1}_\mathbf{A}]_{i,j} = 0$, if $a_{i,j}=0$. The matrix $\mathbf{I}_{N \times N}$ is the $N \times N$ identity matrix. The notations $(\cdot)^\top$,  $\odot$, $\| \cdot \|$,  $\mbox{tr} \{ \cdot \}$, $\mathbb{P} (\cdot)$, $\E \{ \cdot \}$, and $\var( \cdot )$ stand for the transpose, Hadamard product, Euclidian norm of a vector, trace of a matrix, probability, mathematical expectation, and variance, respectively. The notation for diagonal elements of a matrix $\mathbf{A}$ is $\diag(\mathbf{A})$, and the notation $N( 0, \sigma^2)$ stands for Gaussian zero-mean distribution with variance $\sigma^2$.

\subsection{Paper Organization}
In Section~II, we start by recalling graph models and establishing a result, which brings an analogy between Gaussian distribution and Erd\"{o}s--R\'{e}nyi model. This motivates the special role of Erd\"{o}s--R\'{e}nyi model in graph error models which are introduced for different types of graphs. Section~III is devoted to defining and studying graph autocorrelation. In Section~IV, the effect of adjacency matrix mismatch for several GSP applications is studied using simulations and real data examples. Section~V concludes the paper.

\section{GRAPH ERROR MODELS}
\label{sec:graph}

\subsection{Basic Building Blocks}
Let us first go through the basics of graphs, and define the graph models which will be used in the paper.    

Let $\mathcal{G} = (\mathcal{N},\mathcal{E})$ be a directed graph that represents the basis of a graph signal, where $\mathcal{N}$ is the set of $N$ nodes and $\mathcal{E}$ is the set of edges. The adjacency matrix of the graph $\mathcal{G}$, denoted as $\mathbf{A}$, is a matrix that satisfies the conditions $a_{i,i}= 0$ for $i=1,\dots,N$ and $a_{i,j} \neq 0$ if and only if $(i,j) \in \mathcal{E}$, i.e., there is an edge from node $j$ to node $i$.

For developing our graph error models, we will use the Erd\"{o}s--R\'{e}nyi model according to which a random graph is constructed by connecting nodes randomly with a constant probability \cite{ErdosRenyi1959}. The corresponding graph is denoted as $\mathcal{G} = (\mathcal{N}, \epsilon)$ and its adjacency matrix $\mathbf{\Delta}_\epsilon$ is a random $N \times N$ matrix such that $\mathbb{P} ( [\mathbf{\Delta}_\epsilon]_{i,j} = 1 ) = \epsilon$ and $\mathbb{P} ( [\mathbf{\Delta}_\epsilon]_{i,j} = 0 ) = 1 - \epsilon$ for all $i \neq j$, and $[\mathbf{\Delta}_\epsilon]_{i,i} = 0$ for $i = 1, \dots, N$, where each element of the matrix is generated independently from the other elements. 

Another graph model which will be used in this paper is the stochastic block model (SBM)~\cite{Hollandetal1983}, where each node belongs to one of $r$ communities and connections within a community are more common than connections between nodes in different communities. Let $N_1,\dots,N_r$ denote the sizes of the communities and let us assume that the nodes are ordered so that the $i$th node belongs to the $j$th community if and only if $\sum_{l=1}^{j-1}N_l<i\leq \sum_{l=1}^j N_l$. The probability of the edge existence from a node of the $j$th community to a node of the $i$th community is denoted as $p_{i,j}$. 
A special case of the stochastic block model when $p_{i,i}=p_{j,j}=p$ for all $i$ and $j$, and $p_{i,j}=p_{k,l}=q$ for any $i\neq j$ and $k\neq l$, is called planted partition model (PPM).  

Because of the errors in the adjacency matrix learning, the estimated adjacency matrix deviates from the true one. In the following subsections, we introduce graph error models in the form aligned with the objective to study the effect of the errors in the adjacency matrix. All models share the same additive structure, i.e., have the form 
\begin{equation} \label{GenModelW}
\mathbf{W} = \mathbf{A} + \mathbf{E},
\end{equation}
where $\mathbf{W}$ presents the estimated adjacency matrix, $\mathbf{A}$ is the correct adjacency matrix, and $\mathbf{E}$ is an unknown error matrix which can be viewed as an analog of the additive error/distortion/noise component (that applies not to the signal points, but to the signal structure) in the traditional signal processing and time series analysis. Notice that $\mathbf{E}$ needs to be a function of $\mathbf{A}$ because it depends on $\mathbf{A}$ whether distorting an edge means adding or deleting it.
Moreover, the Erd\"{o}s--R\'{e}nyi graph is the basic GSP error/distortion/noise model analogous to the basic Gaussian noise model in the traditional signal processing as will be shown in the next subsection.

The correct adjacency matrix for a given graph signal and a GSP method can be defined as the one which yields the best results. In the case when the data generating process is known, the optimal adjacency matrix in that sense can often be found. On the other hand, when the optimal adjacency matrix is not known, we might still want to use the graph error models to see how much the results vary when the adjacency matrix is perturbed. Then the role of correct adjacency matrix $\mathbf{A}$ can be assigned to an estimated adjacency matrix as well.

\subsection{Motivation for Erd\"{o}s-R\'{e}nyi Graph Error Models}

In the context of our paper, an important characteristic of the Erd\"{o}s-R\'{e}nyi graph is that it does not allow for formation of communities~\cite{MorencyLeus2017}, and if applied on the top of another graph, it will not change the essential structure of that graph. Instead, it just disturbs the spectrum of the original graph as will be shown later in this section. 

The simplicity and the above-mentioned properties are the main reasons for the use of Erd\"{o}s-R\'{e}nyi model as the main block for modelling the mismatches in the graphical structures of graph signals. However, we will  further justify it by bringing an analogy to the basic nature of the Gaussian additive noise for modelling errors related to imperfect fit between signal models and actual measured signal values. 

Similar to the fact that the key for modelling additive noise is the noise probability density function, the key for modelling graph errors will be the graph function or {\it graphon}. The term graphon was introduced in \cite{LovaszSzegedy2006}, in the context of graph sequence theory, and it refers to measurable and symmetric functions mapping from $[0,1]^2$ to $[0,1]$. Graphons have been used, for example, as graph limit objects~\cite{LovaszSzegedy2006, Lovasz2012}, and as building blocks in the kernel graph model~\cite{Bollobasetal2007}. Moreover, they have been used for spectral analysis of the adjacency matrix~\cite{MorencyLeus2017}. See Appendix~A for a brief introduction to graphons.

In traditional signal processing, the main reason for the use of Gaussian random variables as error terms meant to model the discrepancies between the analytic signal model and actual measurements is the central limit theorem (CLT). This theorem states that the distribution of sum of independent and identically distributed random variables tends to the Gaussian distribution as the number of summands grows. With respect to modelling errors in graphs, the Erd\"{o}s-R\'{e}nyi graph plays a special role due to its simplicity. However, the importance of the Erd\"{o}s-R\'{e}nyi graph also follows from the fact that similar to the Gaussian distribution for measurement error modelling, the Erd\"{o}s-R\'{e}nyi graphon leads to a theorem that can be considered as an analog of CLT in the case of modelling the graph structure errors.

Let $\mathcal{W}_\epsilon$ denote the set of all graphons $W$, which is a weighted graph with uncountable set of nodes, satisfying 
\[
\int_0^1\int_0^1 W(x,y)dxdy=\epsilon
\]
where $W(x,y)$ is the edge weight between nodes $x$ and $y$ in the graphon $W$ (see Appendix~A for more details), and $0<\epsilon<1$. Then the following theorem is in order.  

\begin{theorem}
	\label{theorem1}
	Let $W_1,\dots,W_M$ be random samples from $\mathcal{W}_\epsilon$ and $c_1,\dots,c_M$ be random samples from a distribution with support on $[0,1]$ and expected value $c$. Assume that $W_1,\dots,W_M,c_1,\dots,c_M$ are mutually independent. Then for all $0\leq x,y\leq 1$, we almost surely have 
	\[
	\bar{W}(x,y)=\frac{1}{M}\sum_{i=1}^M c_iW_i(x,y)\to c\epsilon \quad \text{as} \quad M\to \infty .
	\]
\end{theorem}
The proof of the theorem is given at the end of Appendix~A. 

Theorem~\ref{theorem1} states that the average of many independent graphons converges to a constant, i.e., to the Erd\"{o}s-R\'{e}nyi graphon. Thus, the theorem further motivates the use of Erd\"{o}s-R\'{e}nyi graphs as the main building block in developing graph error models for GSP, when the summands are considered as noise factors which impede the estimation of the graph structure. In this sense, Theorem~1 is an analog of the CLT.  

\subsection{A Basic Model for Unweighted Graphs}
We start by considering unweighted graphs, for which the adjacency matrix becomes $\mathbf{A} = \mathbf{1}_{\mathbf{A}}$. First, making a somewhat idealistic assumption that the outcome of the graph signal adjacency matrix learning is accurate enough, that is, assuming 
that incorrect graph edge learning is equally probable for any edge in the graph, the learning errors can be simply modelled by the Erd\"{o}s-R\'{e}nyi\footnote{Even though the results in \cite{ErdosRenyi1959} are for undirected graphs only and often Erd\"{o}s-R\'{e}nyi model is considered to imply undirected graph, we will use the term Erd\"{o}s-R\'{e}nyi for both undirected and directed graphs.} graph. 
Then the actually available learned adjacency matrix of a graph signal can be modelled in terms of the following inaccurate version of $\mathbf{A}$
\begin{equation}
\mathbf{W} = \mathbf{A} + \mathbf{\Delta}_\epsilon \odot (\mathbf{1}_{N\times N} - 2 \mathbf{A}). \tag{M1} \label{model1}
\end{equation}  

According to \eqref{model1}, the true adjacency matrix of a graph signal is distorted because of the imperfect learning, and this distortion follows the  Erd\"{o}s-R\'{e}nyi model, where the level of distortion depends on a single parameter, which is the probability $\epsilon$. As a result of the distortion captured by model \eqref{model1}, an edge can be added with probability $\epsilon$, when there is no edge in the true graph, or an edge of the true graph can be removed with the same probability, if there exists an edge in the true graph. It corresponds to flipping the value from 0 to 1 or from 1 to 0 in the adjacency matrix $\mathbf{1}_{\mathbf{A}}$ in the positions corresponding to value 1 in the Erd\"{o}s-R\'{e}nyi adjacency matrix $\mathbf{\Delta}_\epsilon$. This basic model is introduced due to its simplicity as it has a single parameter, $\epsilon$. However, it may not be sufficiently flexible to describe signal graph errors that appear as outcome in any practical graph learning process. For example, it will change the number of edges significantly if $\epsilon$ is large and the number of edges is small. 

Model \eqref{model1} can be easily modified/revised for applying to undirected graphs by defining lower triangular matrices $\mathbf{\Delta}_\epsilon^l$ analogously to $\mathbf{\Delta}_\epsilon$, and then replacing $\mathbf{\Delta}_\epsilon$ by $\mathbf{\Delta}_\epsilon^l+(\mathbf{\Delta}_\epsilon^l)^\top$. It is also worth noting that all the following graph error models will be given for directed graphs and the same technique as above can be used to derive the corresponding model for undirected graphs.

\subsection{Different Probabilities of Missed and Mislearned Edges}
Basic model \eqref{model1} can be easily extended to the case where the probability of removing an edge as a result of distortion from a graph which correctly captures a graph signal, denoted as $\epsilon_1$, is not the same as the probability of adding an edge, which does not exist in the true graph, denoted as $\epsilon_2$. The corresponding inaccurately learned adjacency matrix of a graph signal can then be modelled as
\begin{equation}
\mathbf{W} = \mathbf{A} - \mathbf{\Delta}_{\epsilon_1} \odot \mathbf{A} + \mathbf{\Delta}_{\epsilon_2}\odot (\mathbf{1}_{N\times N}-\mathbf{A}). \tag{M2} \label{model2} 
\end{equation}
The second term on the right-hand side in \eqref{model2} corresponds to edge removal and the third term corresponds to edge addition.
This is a straightforward extension of basic model \eqref{model1}, which is, however, very useful for modelling the typical situation in the graph structure learning when particular algorithms are more or less likely to miss an actually existing edge rather than to mislearn an actully non-existing edge \cite{Dongetal2016}. Model \eqref{model2} can be interpreted as an application of two Erd\"{o}s--R\'{e}nyi graphs on the top of the true graph, where one Erd\"{o}s--R\'{e}nyi graph $\mathcal{G} = (\mathcal{N}, \epsilon_2)$ can only add edges which do not exist in the true graph, while the other Erd\"{o}s-R\'{e}nyi graph $\mathcal{G} = (\mathcal{N}, \epsilon_1)$ can only erroneously remove actually existing edges. It is easy to see that model \eqref{model2} is equivalent to model \eqref{model1} when $\epsilon_1=\epsilon_2=\epsilon$.

To analyze the perturbation that the error above proposed models produce, let us start with a trivial observation that if $\mathbf{A}$ is given by an Erd\"{o}s-R\'{e}nyi graph with probability parameter $\alpha$, then $\mathbf{W}$ from model~\eqref{model2} is also an Erd\"{o}s-R\'{e}nyi graph, and the parameter takes value $\alpha(1-\epsilon_1) + (1-\alpha) \epsilon_2$. 

Moreover, if $\mathbf{A}$ is given by the SBM, after applying graph error model \eqref{model2}, the SBM and the communities in the SBM remain the same as is shown in the following proposition. 
\begin{proposition}
	\label{prop1}
	Assume that $\mathbf{A}$ follows the SBM with probability $p_{i,j}^A$ for edges from a node in the $j$th community to a node in the $i$th community. Then $\mathbf{W}$ from graph error model~\eqref{model2} follows SBM with the same communities as in $\mathbf{A}$ and the probabilities are given by 
	\[
	p_{i,j}^W=p_{i,j}^A(1-\epsilon_1)+(1-p_{i,j}^A)\epsilon_2 = (1-\epsilon_1-\epsilon_2)p_{i,j}^A+\epsilon_2.
	\]
\end{proposition}
{\it Proof:} For $\mathbf{W}$ given by \eqref{model2}, the probability $p_{i,j}^W$ for each edge from the $j$th community to the $i$th community is given by the probability that ``the edge is in $\mathbf{A}$ $(p_{i,j}^A)$ and not removed'', i.e., the probability $(1-\epsilon_1)$, plus the probability that ``the edge is not in $\mathbf{A}$ $(1 - p_{i,j}^A)$ and is added,'' i.e., the probability $(\epsilon_2)$. Since $\mathbf{\Delta}_{\epsilon_1}$ and $\mathbf{\Delta}_{\epsilon_2}$ in \eqref{model2} are independent from $\mathbf{A}$, the two probabilities above are products of their component probabilities. This observation gives the stated result and completes the proof. $\square$

The spectrum of the adjacency matrix can be found in closed form for simple models such as the undirected (for the error model of undirected graphs, see the end of the next subsection) PPM with equal community sizes~\cite{Avrachenkovetal2015}. Thus, we also have the following proposition for the spectrum perturbation because of the graph error given by \eqref{model2}.  
\begin{proposition}
	\label{prop3}
	Assume that $\mathbf{A}$ follows the PPM with $M$ equally large communities, and let $p$ be the probability for edges within communities and $q$ be the probability for edges between communities. If $\mathbf{W}$ follows graph error model~\eqref{model2}, the expected difference between the largest eigenvalues of the true adjacency matrix $\mathbf{A}$ and the mismatched one $\mathbf{W}$ is given by 
	\[
	\E\{\lambda_1(\mathbf{A})-\lambda_1(\mathbf{W})\} = N\frac{(\epsilon_1+\epsilon_2)(p+(M-1)q)-M\epsilon_2}{M} .
	\]
	Moreover, the expected difference between the second to $M$th largest eigenvalues for $\mathbf{A}$ and $\mathbf{W}$, are mutually equal and  given by 
	\[
	\E\{\lambda_k(\mathbf{A})-\lambda_k(\mathbf{W})\}=N\frac{(\epsilon_1+\epsilon_2)(p-q)}{M},\ \ k=2,\dots,M.
	\]
\end{proposition}
{\it Proof:}  
For true adjacency matrix $\mathbf{A}$, the expected value of the largest eigenvalue is $N(p+(M-1)q)/M$ and the expected values of next $M-1$ largest eigenvalues are $N(p-q)/M$ \cite{Avrachenkovetal2015}. From Proposition~\ref{prop1}, we see that the expected value of the largest eigenvalue of $\mathbf{W}$ becomes 
\begin{align*}
\E\{\lambda_1(\mathbf{W})\} = \frac{N}{M} ((1-\epsilon_1 - \epsilon_2) p + \epsilon_2 + (M-1) ((1 - \epsilon_1 - \epsilon_2) q + \epsilon_2))
\end{align*} 
and the the expected values of next $M-1$ largest eigenvalues become 
\begin{align*}
\E\{\lambda_2 (\mathbf{W})\} & = \cdots = \E\{\lambda_{M-1} (\mathbf{W})\} = \E\{\lambda_M (\mathbf{W})\} \\ &= \frac{N}{M} ((1-\epsilon_1-\epsilon_2) p + \epsilon_2 - (1-\epsilon_1 - \epsilon_2) q - \epsilon_2) \\ 
&= N ((1 - \epsilon_1 - \epsilon_2) (p-q)).
\end{align*}
Combining these two results together, the difference of the corresponding expected values can be found to be as it is stated in the the proposition. It completes the proof.  $\square$ 

The relative change of the largest eigenvalue is thus $\epsilon_1+(1-M/(p+(M-1)q))\epsilon_2$ and the relative change of next $M-1$ largest eigenvalues is $\epsilon_1+\epsilon_2$.  

\subsection{Generalized Graph Error Model}
\label{sec:general}
It is assumed in models \eqref{model1} and \eqref{model2} that mislearning the edge status is equally probable (although the probabilities of adding non-existing edge and removing existing edge may be different) for all pairs of nodes. For some adjacency matrix estimation methods, this assumption might not hold strictly even if the pairwise connections were equally strong. To allow the differences in learning accuracy, caused by the structure of the graph or the connectivity strength differences, we formulate the following generalized graph error model.

Consider an unweighted adjacency matrix $\mathbf{A}$. Define $\mathcal{D}=\{\mathbf{D}_1,\dots,\mathbf{D}_K\}$, where $\mathbf{D}_1,\dots,\mathbf{D}_K$ are $N\times N$ matrices satisfying $[\mathbf{D}_k]_{ij}\in \{0,1\}$ for all $k=1,\dots,K$ and $i,j=1,\dots,N$, and $\sum_{k=1}^K\mathbf{D}_k=\mathbf{1}_{N\times N}-\mathbf{I}_{N\times N}$. Each matrix $\mathbf{D}_k$ presents the pairs of nodes, indicated by ones, for which the probabilities of mislearning the existence of the edges are equal. The probabilities of removing edges in the subsets are given in $\bm{\epsilon}_1=\{\epsilon_{11},\dots,\epsilon_{K1}\}$, and the probabilities of adding edges are given in $\bm{\epsilon}_2=\{\epsilon_{12},\dots,\epsilon_{K2}\}$. The graph error model specified by $\mathcal{D}$, $\bm{\epsilon}_1$, and $\bm{\epsilon}_2$ can then be written as 
\begin{align}
\label{model3}
 \mathbf{W} = \mathbf{A}-\sum_{k=1}^K\Delta_{\epsilon_{k1}}\odot \mathbf{D}_k\odot \mathbf{A} + \sum_{k=1}^K \Delta_{\epsilon_{k2}}\odot \mathbf{D}_k\odot (\mathbf{1}_{N\times N}-\mathbf{A}). \tag{M3}
\end{align} 

This construction allows to build very flexible and accurate graph error models, which can adjust to basically any type of graph topology learning errors. However, flexible model requires large $K$, i.e., a lot of parameters. Therefore, it is of interest in practice to only approximate the graph error by using a reasonable value of $K$. The accuracy of such approximation is then appears as trade off with the requirements to the amount and reliability of prior information about the graph structure of a graph signal and the complexity in terms of the value of $K$, i.e., the number of parameters. It is worth noting here that the Erd\"{o}s-R\'{e}nyi graph-based error model \eqref{model2} is a special case of graph error model \eqref{model3}. Indeed, it can be obtained by letting $K=1$ and $\mathbf{D}_1=\mathbf{1}_{N\times N}-\mathbf{I}_{N\times N}$. 

For further insights into the above introduced generalized graph error model~\eqref{model3}, we take as an example the SBM. Let $\mathbf{C}_{k,m}$ denote an $N\times N$ matrix such that $[\mathbf{C}_{k,m}]_{i,j}=1$ if $i\neq j$, $\sum_{l=1}^{k-1}N_l<i\leq \sum_{l=1}^k N_l$ and $\sum_{l=1}^{m-1}N_l<j\leq \sum_{l=1}^m N_l$, and $[\mathbf{C}_{k,m}]_{i,j}=0$, otherwise. Then we can set, for example,  $K=r^2$, $\mathbf{D}_1=\mathbf{C}_{1,1}$, $\mathbf{D}_2=\mathbf{C}_{1,2}, \cdots,\mathbf{D}_{r^2}=\mathbf{C}_{r,r}$, and we obtain a model where the probability of mislearning an edge depends on which communities the start and the end nodes belong to. For such model, the following result about the structure of the true adjacency matrix $\mathbf{A}$ and that of the mismatched one $\mathbf{W}$ as modelled by \eqref{model3} is of interest.

\begin{proposition}
	\label{prop2}
	Assume that the true adjacency matrix $\mathbf{A}$ follows the SBM with $r$ communities, the mismatched adjacency matrix $\mathbf{W}$ follows model~\eqref{model3}, and $\mathbf{D}_1,\dots,\mathbf{D}_K$ are chosen according to the SBM structure of $\mathbf{A}$. Then $\mathbf{W}$ retains SBM with the same communities.  
\end{proposition}

{\it Proof:} When all such pairs of nodes where one node belongs to $i$th community and the other one belongs $j$th community of $\mathbf{A}$ are in the same set $\mathbf{D}_k$, the probability for edges between those pairs of nodes remains constant in the adjacency matrix $\mathbf{W}$ mismatched according to \eqref{model3} as well. The latter trivially means that the SBM structure of $\mathbf{A}$ is retained in $\mathbf{W}$. $\square$

It is also interesting to note that a natural graph error model related to the previously used PPM can be derived using model~\eqref{model3} by letting $K=2$, $\mathbf{D}_1=\sum_{k=1}^r\mathbf{C}_{k,k}$, and $\mathbf{D}_2=\mathbf{1}_{N\times N} -\mathbf{I}_{N\times N}-\sum_{k=1}^r\mathbf{C}_{k,k}$.

\subsection{Models for Weighted Graphs}
The above defined models can be extended to signals on weighted graphs. Then, in addition to adding and removing edges, we have to consider changes in the edge weights and also figure out how to generate the weights for added edges in a mismatched adjacency matrix. 

Let us start with extending graph error model~\eqref{model2} (extension of graph error model~\eqref{model1} is similar). Let $\mathcal{A}$ denote the set of nonzero elements of the true graph adjacency matrix $\mathbf{A}$. The inaccurately learned weighted adjacency matrix can be then modelled as 
\begin{align}
\mathbf{W} = \mathbf{A} + (\mathbf{1}_{N\times N}- \mathbf{\Delta}_{\epsilon_1}) \odot \mathbf{1}_{\mathbf{A}} \odot \mathbf{\Sigma}_c -\mathbf{\Delta}_{\epsilon_1} \odot \mathbf{A} + \mathbf{\Delta}_{\epsilon_2} \odot \mathbf{B} \odot (\mathbf{1}_{N\times N} - \mathbf{1}_{\mathbf{A}}). \tag{M2w} \label{model2w}
\end{align}

Like in graph error model~\eqref{model2}, $\epsilon_1$ is the probability to erroneously removing edges while $\epsilon_2$ is the probability of erroneously adding edges. The second term in \eqref{model2w} perturbs the weights of the remaining edges using an $N\times N$ matrix $\mathbf{\Sigma}_c$ whose elements are drawn from a zero mean Gaussian distribution with variance $c \cdot \sigma^2$, where $\sigma^2$ is the sample variance of $\mathcal{A}$ and $c$ is the variance multiplier parameter. The third term in \eqref{model2w} models the erroneous removal of edges and the fourth term in \eqref{model2w} models the erroneous addition of edges with weights given by the matrix $\mathbf{B}$ which is an $N\times N$ matrix whose elements are derived by the same rules which were used for edge weights of $\mathbf{A}$ if applicable, or alternatively the elements can be drawn from $\mathcal{A}$ with replacement. Typically for practical GSP tasks, only positive edge weights are considered, and then, if the above described process produces negative elements in $\mathbf{W}$, they should be put to zero.

Similarly, the weighted generalized graph error model can be written as a weighted extension of graph error model \eqref{model3}, that is,
\begin{align}
\label{model3w}
\mathbf{W}& = \mathbf{A} + \sum_{k=1}^K\left((\mathbf{1}_{N\times N}-\mathbf{\Delta}_{\epsilon_{k1}}) \odot \mathbf{D}_k \odot \mathbf{1}_{\mathbf{A}} \odot \mathbf{\Sigma}_{ck} \right. \nonumber  \\ 
& \left. -\mathbf{\Delta}_{\epsilon_{k1}} \odot \mathbf{D}_k \odot \mathbf{A} 
+\mathbf{\Delta}_{\epsilon_{k2}} \odot \mathbf{B}_k \odot \mathbf{D}_k \odot (\mathbf{1}_{N\times N} - \mathbf{1}_{\mathbf{A}})\right). \tag{M3w}
\end{align}
where $\epsilon_{k1},\ \epsilon_{k2}$ and $\mathbf{D}_k$, $k=1,\dots,K$ are defined in the same way as in Subsection~\ref{sec:general}, and $\mathbf{B}_k$ and $\mathbf{\Sigma}_{ck}$, $k=1,\dots,K$ are analogous to $\mathbf{B}$ and $\mathbf{\Sigma}_c$ above.

\section{GRAPH AUTOCORRELATION}
In GSP tasks such as filtering or signal recovery from samples, smoothness with respect to the chosen shift matrix is a key assumption. Therefore, smoothness is also instrumental in the shift operator learning from training data, as can be also seen for example in~\cite{Chepurietal2017}. Graph autocorrelation is a measure of smoothness which we will discuss here.\footnote{Later in the paper, we deal with a related concept for multivariate signals in the context of ICA, that is, graph autocorrelation matrix.}

As a model for graph signal we use the moving average model which have been used at least since \cite{Blochletal2010}. Thus, the graph moving average (GMA) signal model of order $m$, denoted hereafter as GMA$(m)$, is an extension for graph signals of the traditional time series moving average (MA) model, and it is given as 
\begin{equation}
\label{GSmodel}
\mathbf{z} = \mathbf{y} + \sum_{l=1}^m \theta_l \mathbf{A}^l \mathbf{y} 
\end{equation}
where $\mathbf{y} \triangleq [y_1, \dots, y_N]^\top$ with $y_1, \dots, y_N \sim N(0, \sigma_y^2)$ being mutually independent Gaussian random variables with zero mean and variance $\sigma_y^2$, and $\theta_1, \dots, \theta_m$ are MA coefficients.
In this paper, we mainly consider the GMA$(1)$ model 
\begin{equation}
\label{GMA1}
\mathbf{z} = \mathbf{y} + \theta_l \mathbf{A} \mathbf{y} = \tilde{\mathbf{A}}\mathbf{y} 
\end{equation}
where $\tilde{\mathbf{A}}=\mathbf{I}_{N\times N}+\theta\mathbf{A}$.

We start with the autocovariance of a graph signal. For a centered graph signal, its autocovariance at lag $k$ with respect to $\mathbf{W}$ is defined as 
\begin{align}
\label{AutoCov1}
s_{\mathbf{z},k} (\mathbf{W}) = \mbox{tr}\{\cov(\mathbf{z}, \mathbf{W}^k\mathbf{z})\}=\E \left\{ \frac{1}{N-k} \mathbf{z}^\top \mathbf{W}^k \mathbf{z} \right\} = \frac{1}{N-k} \E \left\{ \mathbf{z}^\top \mathbf{W}^k \mathbf{z} \right\}. 
\end{align}

Note here that another quadratic form $\mathbf{z}^\top\mathbf{L}\mathbf{z}$, where $\mathbf{L}$ is the graph Laplacian, appears often in GSP literature, for example in~\cite{Chepurietal2017}. Since the off-diagonal elements of the Laplacian matrix are the opposites of those of the adjacency matrix and the diagonal elements of the Laplacian are given by the node degrees whereas the diagonal elements of the adjacency matrix are zeros, the two quadratic forms as measures of smoothness are closely related but not fully equivalent. Large graph autocovariance values indicate smoothness, but in the case of the Laplacian, small values indicate smoothness. Indeed, assume that there are two adjacency matrices $\mathbf{W}_1$ and $\mathbf{W}_2$, and the matrix $\mathbf{W}_1$ gives a larger graph autocovariance value for $\mathbf{z}$ and $k=1$ than that of the matrix $\mathbf{W}_2$. Then for the corresponding Laplacian matrices $\mathbf{L}_1$ and $\mathbf{L}_2$, respectively, it holds that $\mathbf{z}^\top\mathbf{L}_1\mathbf{z}< \mathbf{z}^\top\mathbf{L}_2\mathbf{z}$ any time when autocovariance values are distinct and not close to each other, and only if the autocovariance values are close to each other, then sometimes $\mathbf{z}^\top\mathbf{L}_1\mathbf{z}> \mathbf{z}^\top\mathbf{L}_2\mathbf{z}$. 
 
For a GMA$(1)$ signal $\mathbf{z}$ from~\eqref{GMA1}, we obtain 
\begin{align*}
s_{\mathbf{z},k}(\mathbf{W}) & = \frac{1}{N-k} \E \left\{ \mathbf{y}^\top \tilde{\mathbf{A}}^\top  \mathbf{W}^k \tilde{\mathbf{A}}\mathbf{y}\right\} = \frac{1}{N-k} \mbox{tr} \left\{ \mathbf{W}^k \E\left\{\tilde{\mathbf{A}}\mathbf{y}(\tilde{\mathbf{A}}\mathbf{y})^\top\right\} \right\}  \\
& = \frac{\sigma^2_y}{N-k} \mbox{tr} \left\{\mathbf{W}^k(\mathbf{I}_{N\times N}+ \theta \mathbf{A}^\top) (\mathbf{I}_{N\times N}+\theta \mathbf{A})\right\} 
\end{align*}
where the property $\mbox{tr} \left\{ \mathbf{A} \mathbf{B} \right\} = \mbox{tr} \left\{ \mathbf{B} \mathbf{A} \right\}$ is used in the first step, and the second step follows from the formula of the covariance matrix $\E\left\{\mathbf{z}\mathbf{z}^\top\right\}$ of $\mathbf{z}=\tilde{\mathbf{A}}\mathbf{y}$, see~\cite{Chepuri2017}. Thus, graph autocovariance for GMA$(1)$ signal is a weighted sum of the covariances between the nodes of a graph signal. 

As a standardized version of the graph autocovariance we define graph autocorrelation which is invariant to changes in scales of the graph signal and the matrix $\mathbf{W}$. Then the following definition can be given. 
\begin{definition}
	\label{Autocorr}
	The graph autocorrelation of lag $k$ with respect to matrix $\mathbf{W}$ is  
	\begin{align*} 
	r_{\mathbf{z},k} (\mathbf{W})=\frac{\E \left\{\mathbf{z}^\top \mathbf{W}^k \mathbf{z} \right\}}{\left(\E\{\|\mathbf{z}\|^2\}\E\{\|\mathbf{W}^k\mathbf{z}\|^2\}\right)^{1/2}} .
	\end{align*}
\end{definition}

It is interesting to see how the graph autocorrelation in Definition~\ref{Autocorr} depends on the specific parameters of random graph error model, for example, parameters $\epsilon_1$ and $\epsilon_2$ of graph error model \eqref{model2}. It can be seen that the graph autocorrelation is also a function of the graph adjacency matrix $\mathbf{A}$. Thus, let us also assume that $\mathbf{A} = \mathbf{\Delta}_{\alpha}$, i.e., $\mathbf{A}$ also follows the Erd\"{o}s-R\'{e}nyi model with the probability parameter ${\alpha}$.

For graph autocorrelation $r_{\mathbf{z},1} (\mathbf{W})$ of the GMA(1) signal $\mathbf{z}$, we would like to derive the expected value, when both $\mathbf{A}$ and $\mathbf{W}$ are considered to be random, that is, the expected value $\E_{\mathbf{A},\mathbf{W},\mathbf{z}}\{r_{\mathbf{z},1}(\mathbf{W})\}$ for the lag $k=1$. Even though this expected value is an average over Erd\"{o}s-R\'{e}nyi graphs $\mathbf{A}$, we have verified by extensive simulations that the graph autocorrelation values are very similar for any specific and fixed realization of $\mathbf{A}$. The expected graph autocorrelation value can be then expressed in terms of the essential parameters only, such as the GMA coefficient $\theta$, the probability parameter $\alpha$ in the graph model $\mathbf{A}=\mathbf{\Delta}_{\alpha}$, and for example, the parameters $\epsilon_1$ and $\epsilon_2$ in the graph error model given by \eqref{model2}. The expected value $\E_{\mathbf{A}, \mathbf{W}, \mathbf{z}} \{r_{\mathbf{z},1} (\mathbf{W})\}$ can be used for example to verify, that GMA signal is smoother with respect to its adjacency matrix than the mismatched adjacency matrix. 

\vspace{0.1cm}

{\it Example: Expected value of the graph autocorrelation at $k=1$ for GMA(1) and graph error model \eqref{model2}.} 

\vspace{0.1cm}

We approximate the expected value of the ratio in $r_{\mathbf{z},1}(\mathbf{W})$ by ratio of expected values, and the expected value of product in the denominator by the product of expected values.\footnote{Even though the elements are not independent, it will be seen in simulations that this approximation is accurate.}

We start by rescaling the matrices $\mathbf{A}_a=a\mathbf{A}$ and $\mathbf{W}_w=w\mathbf{W}$ and choosing the variance $\sigma_y^2$ so that the following approximation holds  
\[
\E_{\mathbf{A},\mathbf{W},\mathbf{z}}\{r_{\mathbf{z},1}(\mathbf{W})\} \approx \E_{\mathbf{A}_a,\mathbf{W}_w,\mathbf{z}}\{\mathbf{z}^\top\mathbf{W}\mathbf{z}\}.
\]
In order to have this, $a$, $w$ and $\sigma_y^2$ need to be chosen so that the following three equations are satisfied.
\begin{align}
\label{autocorreqns}
& \frac{1}{N} \E\{\|\mathbf{z}\|^2\} = 1 \nonumber \\
& \frac{1}{N} \E\{\|\mathbf{A}_a\mathbf{z}\|^2\} = 1 \\
& \frac{1}{N} \E\{\|\mathbf{W}_w\mathbf{z}\|^2\} = 1 . \nonumber 
\end{align}

For given values of $N$, $\alpha$, $\epsilon_1$ and $\epsilon_2$, the only unknowns in~\eqref{autocorreqns} are $a$, $w$ and $\sigma_y^2$. Hence we can find $a$, $w$ and $\sigma_y^2$ by simply solving the system of three equations~\eqref{autocorreqns}. Notice that only the third equation includes $w$, which implies that $a$ and $\sigma_y^2$ can be found from the first two equations. However, the expectations in \eqref{autocorreqns} need to be calculated first. For calculating the expected values in~(\ref{autocorreqns}), the fact that $\E\{y_i y_j\} = \sigma_y^2$ if $i=j$ and $\E\{y_i y_j\} = 0$ otherwise can be used together with the assumption that the elements of the matrices $\mathbf{\Delta}_{\alpha}$,  $\mathbf{\Delta}_{\epsilon_1}$, and $\mathbf{\Delta}_{\epsilon_2}$ are independent within each matrix as well as between the matrices.\footnote{For more details, see Appendix~B} 

The parameter $w$ can be found from the last equation in \eqref{autocorreqns} as 
\begin{align*}
w=&\left(\sigma_y^2\left( -\alpha^3N^2\theta^2a^2(\epsilon_1+\epsilon_2-1)^2 \right. \right. + \alpha^2N^2\theta^2a^2(1-\epsilon_1+\epsilon_2(2(\epsilon_1+\epsilon_2)-3)) \\
+&\alpha N^2\theta^2a^2(\epsilon_2-\epsilon_2^2) 
+ \alpha^2N(\theta^2a^2(\epsilon_1+\epsilon_2-1)-(\epsilon_1+\epsilon_2-1)^2) \\
+&\alpha N(1-\epsilon_1+\epsilon_2(2(\epsilon_1+\epsilon_2)-\theta^2a^2-3)) 
+ \left.\left.N(\epsilon_2-\epsilon_2^2)+\alpha(\epsilon_1+\epsilon_2-1)-\epsilon_2 \right)\right)^{-1/2}
\end{align*}
and the parameters $a$ and $\sigma_y^2$ can be found by solving the system of remaining two equations that after substituting the above found $w$ becomes
\[
\left\{
\begin{array}{ll}
\sigma^2_y ((\alpha-\alpha^2)\theta^2a^2N-2\alpha\theta a+1)=1 \\
\sigma^2_y ((\alpha^2-\alpha^3)\theta^2a^4N^2-2\alpha^2\theta a^3N+ (\alpha-\alpha^2)a^2N)=1. \end{array}
\right.
\]

Finally, we find for $\mathbf{W}$ from \eqref{model2} with parameters $\epsilon_1$ and $\epsilon_2$, the closed-form approximate expression for the expected autocorrelation value of the graph signal $\mathbf{z}$ as a function of $\epsilon_1$ and $\epsilon_2$ as 
\begin{align}
\label{DerFin}
\E_{\mathbf{A},\mathbf{W},\mathbf{z}}\{r_{\mathbf{z},1}(\mathbf{W})\} 
\approx \sigma_y^2(\theta \, a \, w \, N (\alpha-\alpha^2) - \alpha \, w)(1-\epsilon_1-\epsilon_2) - w \, \epsilon_2.
\end{align}

The analysis of \eqref{DerFin} is non-trivial because, for example, $w$ depends on all other parameters. Thus, we leave it to a numerical study in the next section.

\section{NUMERICAL STUDY}
\label{sec:simul}

\subsection{Study Case 1: Graph Autocorrelation of GMA(1) Signal for the Mismatched Adjacency Matrix}

In our first example, we examine how the expected graph autocorrelation of the GMA$(1)$ signal derived in \eqref{DerFin} changes when the matrix with respect to which it is computed is gradually changed from the adjacency matrix of the signal according to graph error model \eqref{model2}. We assume that $\mathbf{z}$ is a GMA$(1)$ signal with an unweighted Erd\"{o}s-R\'{e}nyi adjacency matrix $\mathbf{A}$ with probability $\alpha$ and coefficient $\theta$ as parameters. 

\begin{figure}[htb]
		\begin{minipage}[b]{1.0\linewidth}
			\centering
			\centerline{\includegraphics[width=15cm]{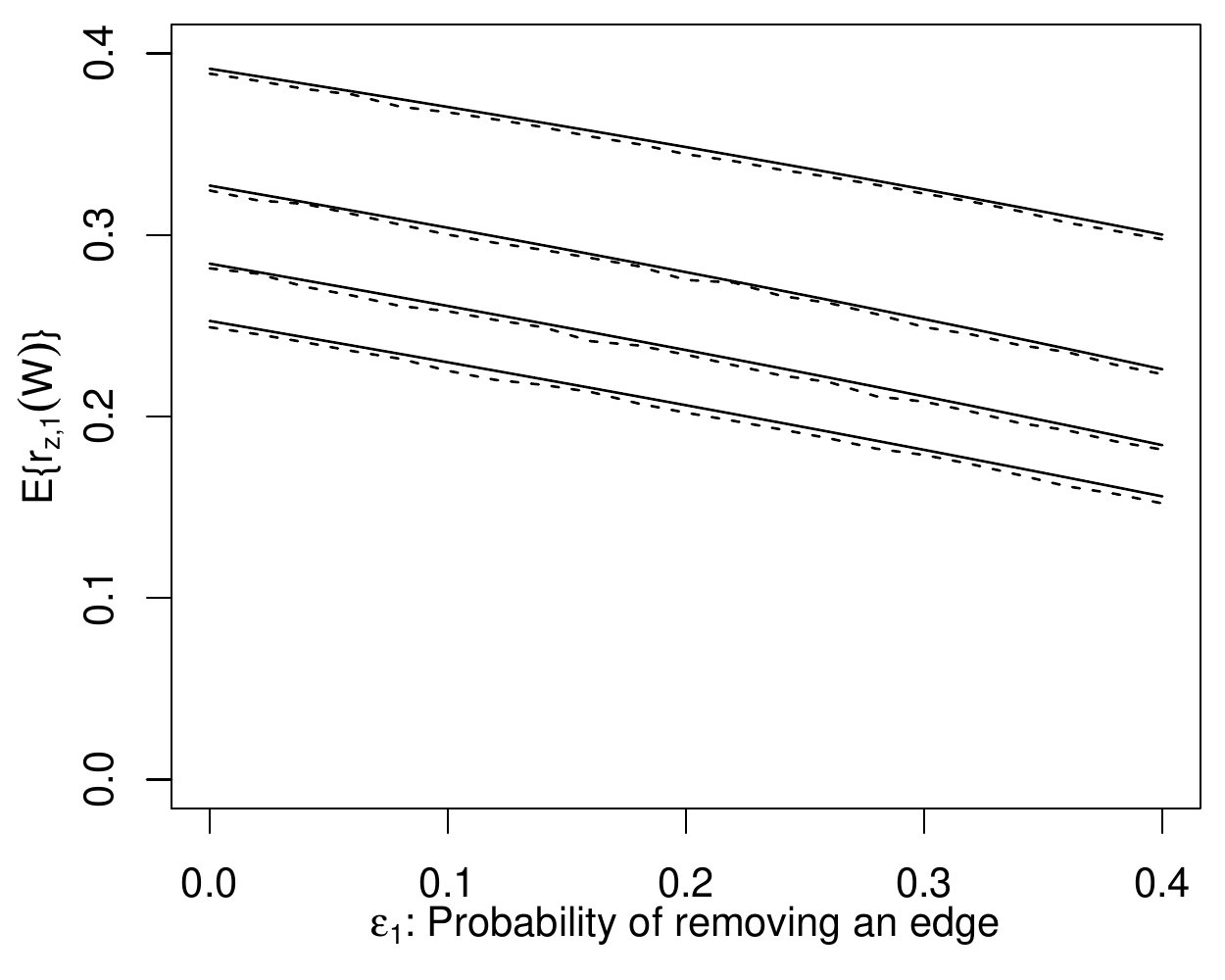}}
		\end{minipage}
		\caption{Theoretical (solid) and numerical (dash) values of graph autocorrelations of order $k=1$. The lines from top to bottom are for $\epsilon_2=0,0.02,0.04,0.06$.}
		\label{fig2}
	
	\end{figure}

\begin{figure}[htb]
		\begin{minipage}[b]{1.0\linewidth}
			\centering
			\centerline{\includegraphics[width=15cm]{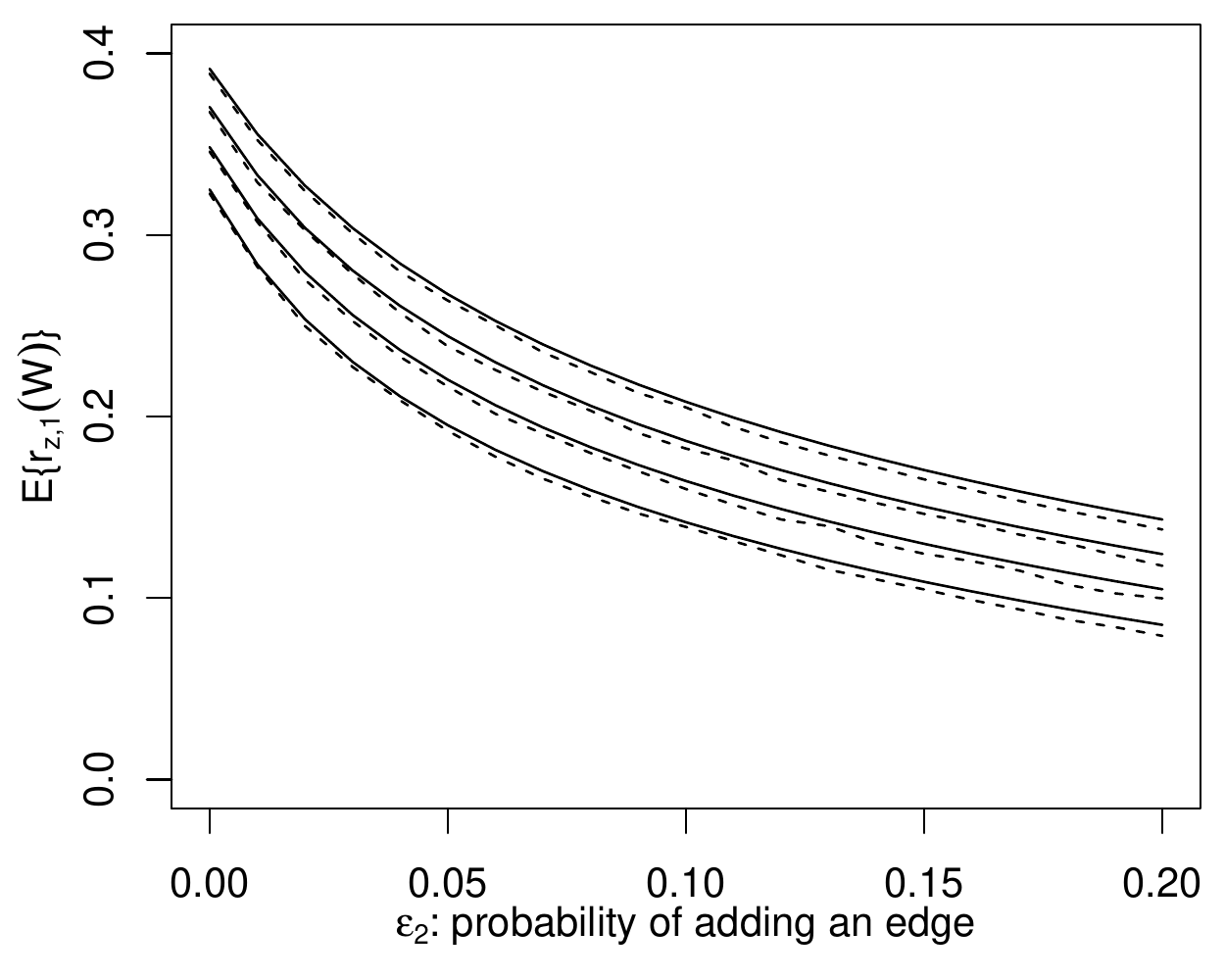}}
		\end{minipage}
		\caption{Theoretical (solid) and simulated (dash) values of graph autocorrelations of order $k=1$. The lines from top to bottom are for $\epsilon_1=0,0.1,0.2,0.3$.}
	   \label{fig3}
	
	\end{figure}

Figs.~\ref{fig2} and~\ref{fig3} illustrate the behavior of the graph autocorrelation as a function of $\epsilon_1$ and $\epsilon_2$, respectively, when one of them varies and the other one is kept fixed. Both theoretical values and averages of 2000 graph autocorrelations from simulated datasets are shown. For each run, a new Erd\"{o}s-R\'{e}nyi adjacency matrix $\mathbf{A}$ is generated with parameters $N=500$ and $\alpha=0.05$, and the GMA coefficient is $\theta=0.5$. The choice of the variance parameter value $\sigma_y^2$ has no effect on the results because of the standardization in the graph autocorrelation (see the definition).

The figures show that the graph autocorrelation is decreasing with respect to both $\epsilon_1$ and $\epsilon_2$, which suggests that graph autocorrelation is useful for estimating the adjacency matrix even if we only have a single realization of the graph signal.
	
\subsection{Study Case 2: Graph Error Effect on GMA Graph Filter} \label{sec:GMAfilter_example}

As in any other GSP task, the adjacency matrix has a key role in graph filtering. In fact, any linear and shift-invariant filter (commutes with other shift-invariant filters) can be written as a polynomial of the form~\cite{SandryhailaMoura2013} 
\begin{align*}
F_{\rm GMA} (\mathbf{W}) = h_0\mathbf{I}_{N\times N}+h_1\mathbf{W}+\dots+h_K\mathbf{W}^K
\end{align*}
where $K$ is smaller than or equal to the degree of the minimal polynomial of $\mathbf{W}$.

The coefficients $h_0,\dots,h_K$ can be found by using least squares for solving the system of equations 
\begin{align*}
h_0+h_1\lambda_1+\dots+h_K\lambda_1^K&=\alpha_1  \\
& \vdots \\
h_0+h_1\lambda_N+\dots+h_K\lambda_N^K&=\alpha_N  
\end{align*} 
where $\lambda_1,\dots,\lambda_N$ are eigenvalues of $\mathbf{W}$ ordered in increasing order of their distance to the maximum magnitude, i.e., $d_n = | \mbox{max} \{ |\lambda_1|, \dots,| \lambda_N|\} - \lambda_n|$. Then $\lambda_1,\dots,\lambda_N$ can be interpreted as graph frequencies from lowest to highest~\cite{SandryhailaMoura2014}, and $\alpha_1, \dots, \alpha_N$ are the corresponding frequency responses. 

In this example, we study the filter sensitivity to the choice of the adjacency matrix, when the high-pass filter is defined by the following frequency responses 
\begin{equation*}
\alpha_n=\left\{\begin{array}{ll} 1,\ \  \mathrm{ if }\ \  d_n>\mbox{median}\{d_1,\dots,d_N\}  \\
0,\ \ \ \mathrm{otherwise}
\end{array}
\right.
\end{equation*}
and it is applied to detecting malfunctioning sensors as in~\cite{SandryhailaMoura2014}. The data are the daily temperatures in the year 2016 from 150 Finnish weather stations~\cite{ClimaticData2016}. Stations that had at most one missing observation were selected, and since those missing observations were from the same day in November, after dropping that day out, we have a clean data of 365 days (leap year). The technique to detect the outlying measurements, presented in~\cite{SandryhailaMoura2014}, is to high-pass filter the data and threshold the Fourier transform coefficients of the output by the maximum absolute value of the GFT coefficients from the three previous days. If any of the coefficients exceed the threshold value, it is diagnosed that at least one of the sensors is not working properly. 

In this example, there is no obvious choice of a benchmark adjacency matrix,\footnote{In other example, there always exists an obvious choice for benchmark adjacency matrix.} but we choose a weighted 6-nearest neighbors graph like in~\cite{SandryhailaMoura2014} with edge weights 
\begin{equation}
\label{weights}
a_{kl}=\frac{e^{-(d_{kl}/20)^2}}{\sqrt{\sum_{j\in\mathcal{N}_k}e^{-(d_{kj}/20)^2}\sum_{j\in\mathcal{N}_l}e^{-(d_{lj}/20)^2}}}
\end{equation}
where $d_{kl}$ is the distance between the locations of the $k$th and $l$th sensors in kilometers. We use the general graph error model because it is not sensible to connect two stations that are far away from each other. We set the threshold to 250 kilometers above which a pair of stations is treated as no longer connected. Otherwise, the probabilities of removing and creating connections have constant values $\epsilon_1$ and $\epsilon_2$ between the pairs. Hence, in graph error model (\ref{model3w}), which is then applicable here, we have the matrices 
\begin{align*}
[\mathbf{D}_1]_{kl}=\left\{\begin{array}{ll} &1,\ \  \mathrm{ if }\ \  d_{kl}\leq 250 \\
& 0,\ \ \ \mathrm{otherwise} \end{array}\right. \\
[\mathbf{D}_2]_{kl}=\left\{\begin{array}{ll} &1,\ \  \mathrm{ if }\ \  d_{kl} > 250 \\
& 0,\ \ \ \mathrm{otherwise} \end{array}\right.
\end{align*}
and probabilities $\epsilon_{11}=\epsilon_1$, $\epsilon_{12}=\epsilon_2$, $\epsilon_{21}=\epsilon_{22}=0$. 

By perturbing the edge weights, we deviate slightly from graph error  model~\eqref{model3w}, so that the variance of the Gaussian component related to incoming edges of a given node is the variance multiplier $c$ times the variance of the incoming edge weights of that node in the 6-nearest neighbors graph, not the variance of the edge weights of the whole 6-nearest neighbors graph. We do not allow negative weights, but use zero weights instead.

In the experiment, we change one sensor value at a time to +20 Celsius degrees and perform the test described above using adjacency matrices given by all combinations of edge removing probabilities $\epsilon_1=0,0.01,0.02,0.03$, edge adding probabilities $\epsilon_2=0,0.1,0.2,0.3$, and variance parameter values $c=0,0.005,0.01,0.015$. 

Finding the malfunctioning sensors can be seen as a binary classification task. Theoretically, all classifiers (the same method with different graphs) have equal proportion of false positives of about $25\%$, because the maximum GFT coefficient today is larger than the maximum of the GFT coefficients from the three previous days one out of four times when there are no outliers. Hence, we can focus on the accuracy in finding the true positives. The benchmark adjacency matrix given by $\epsilon_1=\epsilon_2=c=0$ found the malfunctioning sensors with $84\%$ accuracy. For the other combinations, 200 realizations of the adjacency matrix are generated. 

\begin{figure}[htb]
	\begin{minipage}[b]{1.0\linewidth}
		\centering
		\centerline{\includegraphics[trim={3.2cm 9cm 4.2cm 9.5cm},clip,width=15cm]{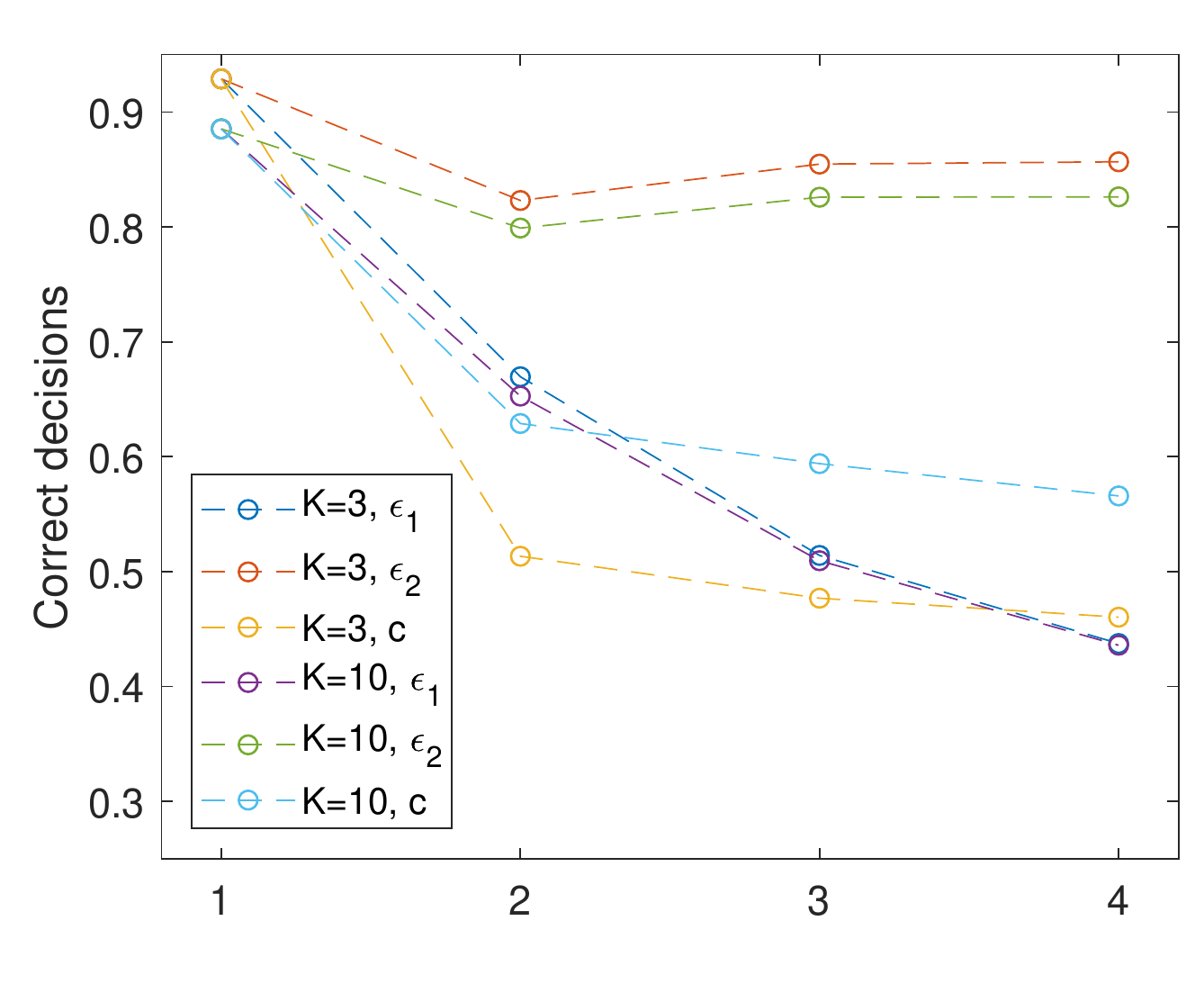}}
	\end{minipage}
	\caption{The share of correctly detected outlying sensor values for $\epsilon_1=0,0.01,0.02,0.03$, $\epsilon_2=0,0.1,0.2,0.3$ and $c=0,0.005,0.01,0.015$.}
	\label{fig4}
	
\end{figure}

In Fig.~\ref{fig4}, we give the averages for different values of parameters of interest when other parameters are zeros. There are no visible patterns of interactions between the parameters, i.e., the shapes of the curves look identical at all levels of the other two parameters. The results show that adding edges to the graph has no effect, but removing edges weakens the performance. Also, the perturbation of the weights has negative impact, though the reduction is significant only for values from 0 to 0.005, and for $c=0.005,0.01,0.015$, the numbers are approximately the same. This suggests that the number of edges in the 6-nearest neighbors graph might be too small for these purposes, but that the edge weights are good as they are, because small changes in them worsen the performance. The reason why addition of edges has smaller effect can be partly explained by the fact that they have smaller edge weight than the edges in the original nearest neighbors graph.

\subsection{Study Case 3: Graph Error Effect on Graph Autoregressive Moving Average  Filter}

First order graph autoregressive moving average (GARMA) filter~\cite{Isufietal2017a} has coefficients $\phi$, $\psi$ and $c$, and uses some graph Laplacian matrix $\mathbf{L}$. We use here the translated normalized Laplacian
\begin{equation*}
\mathbf{L} =  - \mathbf{T}^{-1/2} \mathbf{W} \mathbf{T}^{-1/2}
\end{equation*}  
where $\mathbf{T}$ is a diagonal matrix of node degrees and $\mathbf{W}$ is a symmetric adjacency matrix. Note that the eigenvalues of $\mathbf{L}$ are in the interval $[-1,1]$, but as in the GMA filter case, they can be very different for the cases of the true $\mathbf{A}$ and the estimated $\mathbf{W}$ adjacency matrices.

The output of GARMA(1) filter is given by $\mathbf{z} = \mathbf{y} + c \mathbf{x}$, where $\mathbf{x}$ is the input and $\mathbf{y}$ is the state after convergence of the of recursion $\mathbf{y}_{t+1} = \phi\mathbf{L} \mathbf{y}_t + \psi \mathbf{x}$. GARMA filter of order $K$ can be constructed from $K$ parallel GARMA(1) filters, whose coefficients are designed based on the eigenvalues of $\mathbf{L}$.

Again, the errors in the adjacency matrix $\mathbf{W}$ have a significant effect on the filter directly via $\mathbf{L}$, but also via distorting the spectrum of $\mathbf{L}$. The errors then can reflect even on the GARMA filter stability.  

We are following a simulation setup in~\cite{Isufietal2017b}. Undirected and unweighted graphs are created by generating $N=100$ random points on the area $[0,1]\times [0,1]$ using the uniform distribution, and connecting two points if the distance between them is less than $0.15\sqrt{2}$. 

Let $\mathbf{L}=\mathbf{V}\mathbf{\Lambda}\mathbf{V}^\top$ be the eigendecomposition of the translated normalized graph Laplacian, and let $\lambda_n$ and $\mathbf{v}_n$ denote the $n$th diagonal element of $\mathbf{\Lambda}$ and the $n$th column vector of $\mathbf{V}$, i.e., the $n$th 
eigenvalue and eigenvector, respectively.  The eigenvalues are in the interval $[-1,1]$.  Then the graph signal is given by $\mathbf{x}=\bar{\mathbf{x}}+\mathbf{n}$, where $\bar{\mathbf{x}}$ is a low frequency signal satisfying 
$\bar{\mathbf{x}}^\top\mathbf{v}_n=1$, if $\lambda_n<0$, and $\bar{\mathbf{x}}^\top\mathbf{v}_n=0$, otherwise, and $\mathbf{n}$ is 
Gaussian noise with zero mean and covariance matrix $0.1\mathbf{I}_{N \times N}$.

Frequency responses of GARMA graph filters of orders $K=1,3,5,7$ are designed to match the frequency content of the signal $\bar{\mathbf{x}}$. Outputs of the filters, denoted by $\mathbf{z}^{(e)}$, are compared to the output of the ideal filter, denoted by $\mathbf{z}^{(d)}$. The performance is then measured using the square root of the mean square error 
\begin{equation}
\label{ARMAerror}
\sigma_e=[\mbox{tr} (\mathbf{e}\mathbf{e}^\top )/N]^{1/2} \nonumber
\end{equation} 
where $\mathbf{e}=\mathbf{z}^{(e)}-\mathbf{z}^{(d)}$. 

\begin{figure}[htb]
	\begin{minipage}[b]{1.0\linewidth}
		\centering
		\centerline{\includegraphics[trim={3.2cm 9cm 4.2cm 9.5cm},clip,width=15cm]{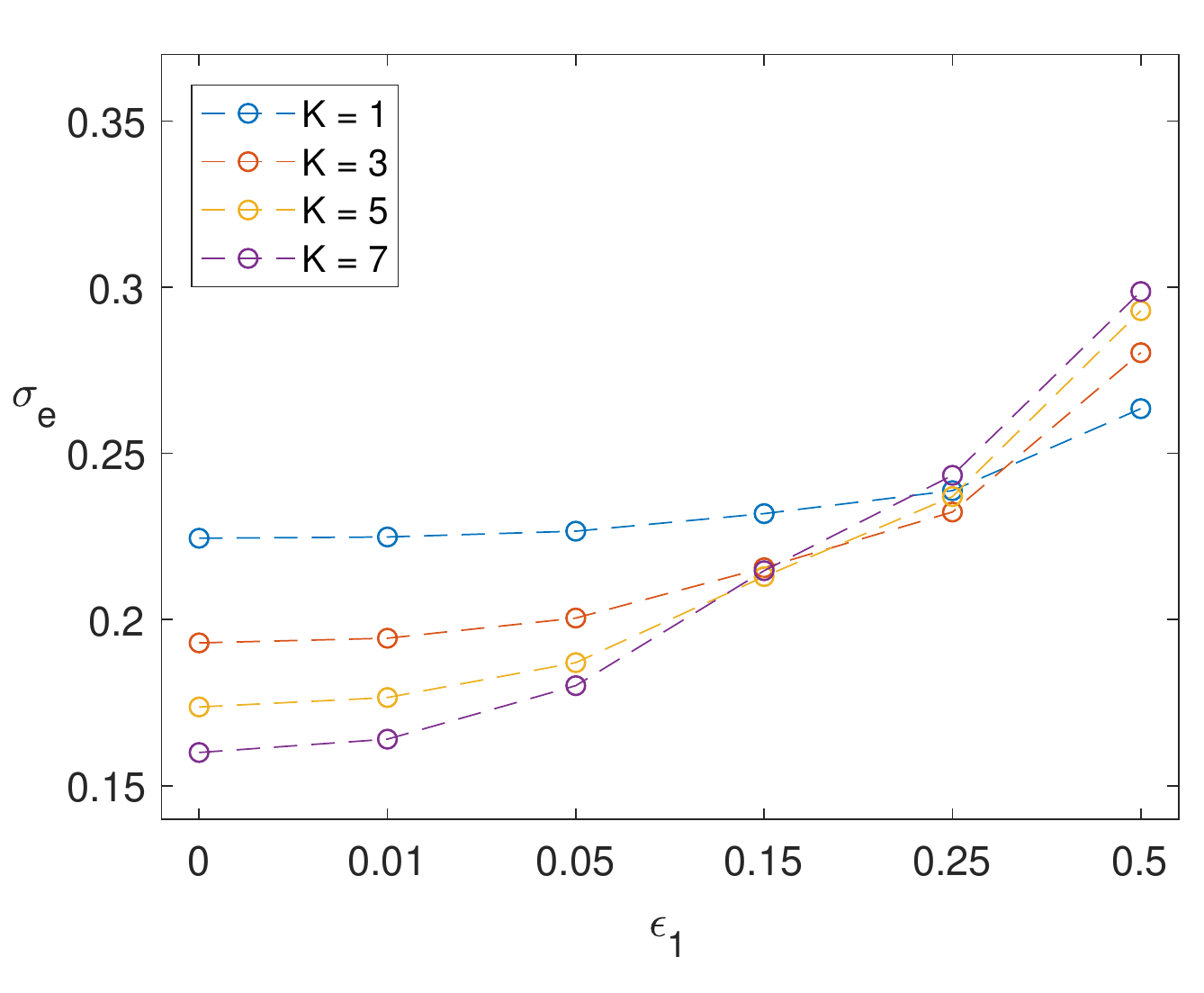}}
	\end{minipage}
	\caption{Averages of $\sigma_e$  over 2000 repetitions for GARMA filter orders $K=1,3,5,7$ and probabilities $\epsilon_1$  in graph error model~\eqref{model2}, when $\epsilon_2=0$.}
	\label{fig5}
\end{figure}

\begin{figure}[htb]
	\begin{minipage}[b]{1.0\linewidth}
		\centering
		\centerline{\includegraphics[trim={3.2cm 9cm 4.2cm 9.5cm},clip,width=15cm]{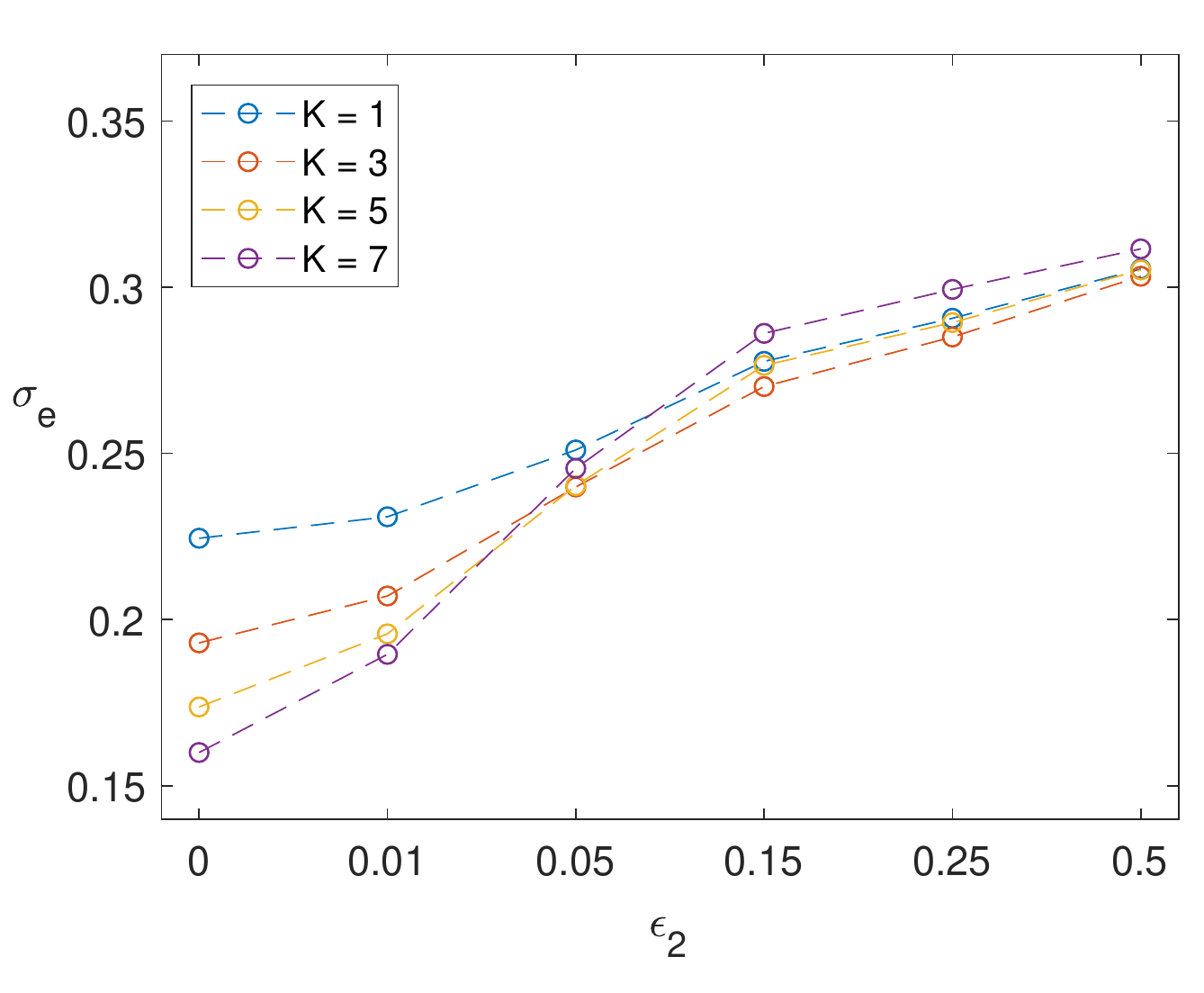}}
	\end{minipage}
	\caption{Averages of $\sigma_e$ over 2000 repetitions for GARMA filter orders $K=1,3,5,7$ and probabilities $\epsilon_2$ in graph error model~\eqref{model2}, when $\epsilon_1=0$.}
	\label{fig6}
\end{figure}

The values in Figs.~\ref{fig5}~and~\ref{fig6} are averages over 2000 runs for each pair $(\epsilon_1,\epsilon_2)$ in graph error model~(\ref{model2}) that is applicable here. The results show that higher order GARMA filters are more accurate when the correct adjacency matrix is used. On the other hand, the lower order filters are more robust to graph errors, and thus all filters perform almost equally when the error probabilities grow. 

\subsection{Study Case 4: Graph Error Effect on ICA of Graph Signals}
\label{sec:ica} 
In this example, we use both numerical study and theoretical results to investigate the effect of non-optimal signal graph knowledge to the performance of GraDe~\cite{Blochletal2010,GraphBSS2020} for graph signal separation in ICA problem.

Let $\mathbf{X} \in \mathbb{R}^{P\times N}$ denote centered $P$-dimensional graph signal generated as a mixture of independent components according to the model \cite{GraphBSS2020} 
\begin{equation}
\mathbf{X} = \mathbf{\Omega} \mathbf{Z} \nonumber 
\end{equation}
where $\mathbf{\Omega} \in \mathbb{R}^{P\times P}$ is a full rank mixing matrix, $\mathbf{Z} \in \mathbb{R}^{P\times N}$ is the matrix of $P$ mutually independent graph signals with zero means and unit variances. 

The goal of ICA is to estimate the unmixing matrix $\mathbf{\Gamma} = \mathbf{\Omega}^{-1}$ using only the signal matrix $\mathbf{X}$.

Let $\mathbf{X}_{\rm w} = \hat{\mathbf{S}}_0^{-1/2} \mathbf{X}$ be the whitened signals, where $\hat{\mathbf{S}}_0$ is the sample covariance matrix of $\mathbf{X}$. In GraDe, the unmixing matrix estimate is obtained by diagonalizing/jointly diagonalizing one or more graph autocorrelation matrices given as 
\begin{equation}
\hat{\mathbf{S}}_k(\mathbf{W}) = \frac{1}{N-k} (\mathbf{X}_{\rm w} \mathbf{W}^k \mathbf{X}_{\rm w}^\top),\ \ k=1,\dots, K \nonumber
\end{equation}
i.e., by finding the orthogonal $\mathbf{U}$ which maximizes the objective function 
\begin{equation}
\sum_{k=1}^K\|\diag(\mathbf{U}\hat{\mathbf{S}}_k(\mathbf{W})\mathbf{U}^\top)\|^2. \nonumber
\end{equation}

The unmixing matrix estimate is then given as $\hat{\mathbf{\Gamma}} = \mathbf{U} \hat{\mathbf{S}}_0^{-1/2}$. A fast algorithm for the joint diagonalization is available in~\cite{Clarkson1988} and it is applicable for the case when the shift matrix $\mathbf{W}$ is chosen to be symmetric, or the graph-autocorrelation matrices are symmetrized. The unmixing matrix estimate for an inaccurately learned adjacency matrix $\mathbf{W}$ is denoted as $\hat{\mathbf{\Gamma}} (\mathbf{W})$. Notice that GraDe reduces to the well-known second-order blind identification (SOBI) estimator~\cite{Belouchranietal1997}, when $\mathbf{W}$ is the cyclic graph.

We will use the following (see~\cite{Miettinenetal2016} for details) asymptotic result, derived in the context of the SOBI estimator, for an unmixing matrix estimate $\hat{\mathbf{\Gamma}}$ obtained using joint diagonalization of matrices $\hat{\mathbf{S}}_1,\dots,\hat{\mathbf{S}}_K$. When $\mathbf{\Omega} = \mathbf{I}_{P \times P}$, for $i\neq j$, we have
\begin{align}
\sqrt{N}\,(\hat{\gamma}_{ii}-1) = -\frac{1}{2} \sqrt{N} \, ([\hat{\mathbf{S}}_0]_{ii}-1) + o_p(1) \nonumber 
\end{align}
and
\begin{align}
\label{GraDeAsymp}
\sqrt{N}\,\hat{\gamma}_{ij} \!=\! 
\frac{\sum_k (\lambda_{ki} - \lambda_{kj}) (\sqrt{N} \, [\hat{\mathbf{S}}_k]_{ij} - \lambda_{ki} \sqrt{N} \, [\hat{\mathbf{S}}_0]_{ij} ) } {\sum_k (\lambda_{ki} - \lambda_{kj})^2} + o_p(1) 
\end{align}
where $\lambda_{ki} \triangleq \E \{[\mathbf{S}_k]_{ii} \}$, and $o_p(1)$ stands for negligible terms. The diagonal elements of $\hat{\mathbf{\Gamma}}$ do not depend asymptotically on $\hat{\mathbf{S}}_1, \dots, \hat{\mathbf{S}}_K$, and thus, in the case of graph signals ICA, do not depend on $\mathbf{W}$. Therefore, the sum of variances of the off-diagonal elements
\begin{align} \label{SOV}
\mbox{SOV}(\hat{\mathbf{\Gamma}}(\mathbf{W})) = N\sum_{j\neq i} \var( \hat{\mathbf{\Gamma}} (\mathbf{W})_{ij})
\end{align}
can be used when comparing the separation efficiencies induced by different choices of $\mathbf{W}$. We will use the ratio of the sums given as 
\[ R(\mathbf{W}_1,\mathbf{W}_2) = \frac{\mbox{SOV} (\hat{\mathbf{\Gamma}} (\mathbf{W}_1))}{\mbox{SOV} (\hat{\mathbf{\Gamma}} (\mathbf{W}_2))} .
\] 
Particularly, $\mathbf{W_1}$ is the adjacency matrix which is used for generating the source components and $\mathbf{W}_2$ is a perturbed version of the adjacency matrix. Using the ratio of the sum of variances instead of the sums helps to visualize the efficiency losses.

We consider ICA model where the independent components are GMA(1) signals with symmetric and unweighted adjacency matrix $\mathbf{A}$, and evaluate the performance of the GraDe estimate with $K=1$ and the matrix $\mathbf{W}$, which is a mismatched version $\mathbf{A}$ obtained according to graph error model~\eqref{model2} for undirected graphs. In this model, the asymptotic variances of $\hat{\mathbf{\Gamma}} (\mathbf{W}))$, which are needed for computing $\mbox{SOV}( \hat{\mathbf{\Gamma}} (\mathbf{W}))$, can be calculated using~\eqref{GraDeAsymp}. Expressions for these asymptotic variances can be derived in the same way as the corresponding formulas for the variances in the time series context in \cite{Miettinenetal2016}.\footnote{Indeed, the extension of asymptotic variance expression in \cite{Miettinenetal2016} from the time series context to the context of graph signals is straightforward, but the formula appears to be too long to be presented here because of the space limitation.}

The performance in the numerical simulations is measured using the minimum distance (MD) index~\cite{Ilmonenetal2010}
\begin{equation}
D(\hat{\mathbf{\Gamma}}) \triangleq \frac{1}{\sqrt{P-1}}\inf_{\mathbf{C}\in
	\mathcal{C}}\|\mathbf{C}\hat{\mathbf{\Gamma}}\mathbf{\Omega}-\mathbf{I}_{P \times P}\| \nonumber 
\end{equation}
where $\mathcal{C} \triangleq \{ \mathbf{C}\ : \mbox{ each row and column of $\mathbf{C}$ has exactly one}$ $\mbox{non-zero element} \}$.
The MD index takes values between zero and one, and it is invariant with respect to the mixing matrix.
Also, there is a connection between the minimum distance index and the sum of variances of the off-diagonal elements when  $\mathbf{\Omega}=\mathbf{I}_{P \times P}$, given as 
\begin{equation}
\label{EMD}
N(P-1)\E\{D(\hat{\mathbf{\Gamma}})^2\}\to \mbox{SOV}(\hat{\mathbf{\Gamma}}),\ \mathrm{as}\  N\to \infty
\end{equation}
where SOV is defined in~\eqref{SOV}. 

For two sets of estimates, $\mathbf{W}_1$ and $\mathbf{W}_2$, we define 
\[
\hat{R}(\mathbf{W}_1,\mathbf{W}_2)=\frac{\ave\{D(\hat{\mathbf{\Gamma}} (\mathbf{W}_1))^2\}}{\ave\{D(\hat{\mathbf{\Gamma}}(\mathbf{W}_2))^2\}}
\]
where the averages are found over 1000 Monte Carlo trials.
Equation~\eqref{EMD} implies that $\hat{R}(\mathbf{W}_1,\mathbf{W}_2)\approx R(\mathbf{W}_1,\mathbf{W}_2)$ for large $N$. 

\begin{table}[ht]
	\centering
	\caption{$R(\mathbf{A},\mathbf{W})$ for $\mathbf{A}$ with $\alpha=0.05$ and $\mathbf{W}$ given by $\epsilon_1=0,0.1,\dots,0.5$ and $\epsilon_2=0,0.01,\dots,0.05$.}
	\begin{tabular}{rrrrrrr}
		\hline
		$\epsilon_1\backslash \epsilon_2$ & 0 & 0.01 & 0.02 & 0.03 & 0.04 & 0.05 \\ 
		\hline
		0 & 1.00 & 0.81 & 0.68 & 0.58 & 0.51 & 0.45 \\ 
		0.1 & 0.88 & 0.71 & 0.58 & 0.49 & 0.43 & 0.38 \\ 
		0.2 & 0.77 & 0.60 & 0.49 & 0.41 & 0.35 & 0.30 \\ 
		0.3 & 0.66 & 0.50 & 0.40 & 0.33 & 0.28 & 0.24 \\ 
		0.4 & 0.56 & 0.41 & 0.31 & 0.26 & 0.21 & 0.18 \\  
		0.5 & 0.46 & 0.31 & 0.24 & 0.19 & 0.15 & 0.13 \\ 
		\hline
	\end{tabular}
	\label{tab:theor}
\end{table}

\begin{table}[ht]
	\centering
	\caption{$\hat{R}(\mathbf{A},\mathbf{W})$ from 1000 repetitions for $\alpha=0.05$ and $\epsilon_1=0,0.1,\dots,0.5$ and $\epsilon_2=0,0.01,\dots,0.05$.}
	\begin{tabular}{rrrrrrr}
		\hline
		$\epsilon_1\backslash \epsilon_2$ & 0 & 0.01 & 0.02 & 0.03 & 0.04 & 0.05 \\
		\hline
		0 & 1.00 & 0.81 & 0.67 & 0.56 & 0.48 & 0.44 \\ 
		0.1 & 0.88 & 0.65 & 0.56 & 0.47 & 0.43 & 0.35 \\ 
		0.2 & 0.76 & 0.59 & 0.46 & 0.40 & 0.34 & 0.29 \\ 
		0.3 & 0.62 & 0.46 & 0.37 & 0.32 & 0.27 & 0.25 \\ 
		0.4 & 0.52 & 0.37 & 0.29 & 0.25 & 0.21 & 0.20 \\ 
		0.5 & 0.43 & 0.31 & 0.25 & 0.20 & 0.17 & 0.16 \\ 
		\hline
	\end{tabular}
	\label{tab:simul}
\end{table}

Erd\"{o}s--R\'{e}nyi matrices $\mathbf{A}=\mathbf{\Delta}_\alpha$ with different values of $\alpha$ are used as the adjacency matrices of GMA signals. The estimate $\hat{\mathbf{\Gamma}}(\mathbf{A})$ (with true $\mathbf{A}$) is a natural benchmark to which we compare the estimates obtained using $\mathbf{W}$. 

In Tables~\ref{tab:theor} and~\ref{tab:simul}, the values of $R(\mathbf{A},\mathbf{W})$ and $\hat{R}(\mathbf{A},\mathbf{W})$ are shown, respectively, when $\mathbf{A}$ is $1000\times 1000$ matrix with $\alpha=0.05$ and there are $p=4$ independent components generated using \eqref{GSmodel} with $\theta=0,\ 0.2,\ 0.4$, and $0.6$. For Table~\ref{tab:simul}, we generate 1000 datasets for each pair $(\epsilon_1,\epsilon_2)$ and always generate a new $\mathbf{W}$. In Table~\ref{tab:theor}, the sum of variances is an average for ten $\mathbf{W}$'s, even though $\mbox{SOV} (\hat{\mathbf{\Gamma}} (\mathbf{W}))$ is quite stable for fixed $\epsilon_1$ and $\epsilon_2$. The simulation results match the theoretical values quite well. When looking at the results with respect to error probabilities $\epsilon_1$ and $\epsilon_2$, it seems that GraDe is more sensitive to adding irrelevant edges than missing the real edges. However, notice that when $\alpha=0.05$, then in numbers of mislearned edges $\epsilon_2=0.01$ corresponds to $\epsilon_1=0.19$.

\begin{figure}[htb]
	\begin{minipage}[b]{1.0\linewidth}
		\centering
		\centerline{\includegraphics[trim={0cm 0cm 8.5cm 17.5cm},clip,width=15cm]{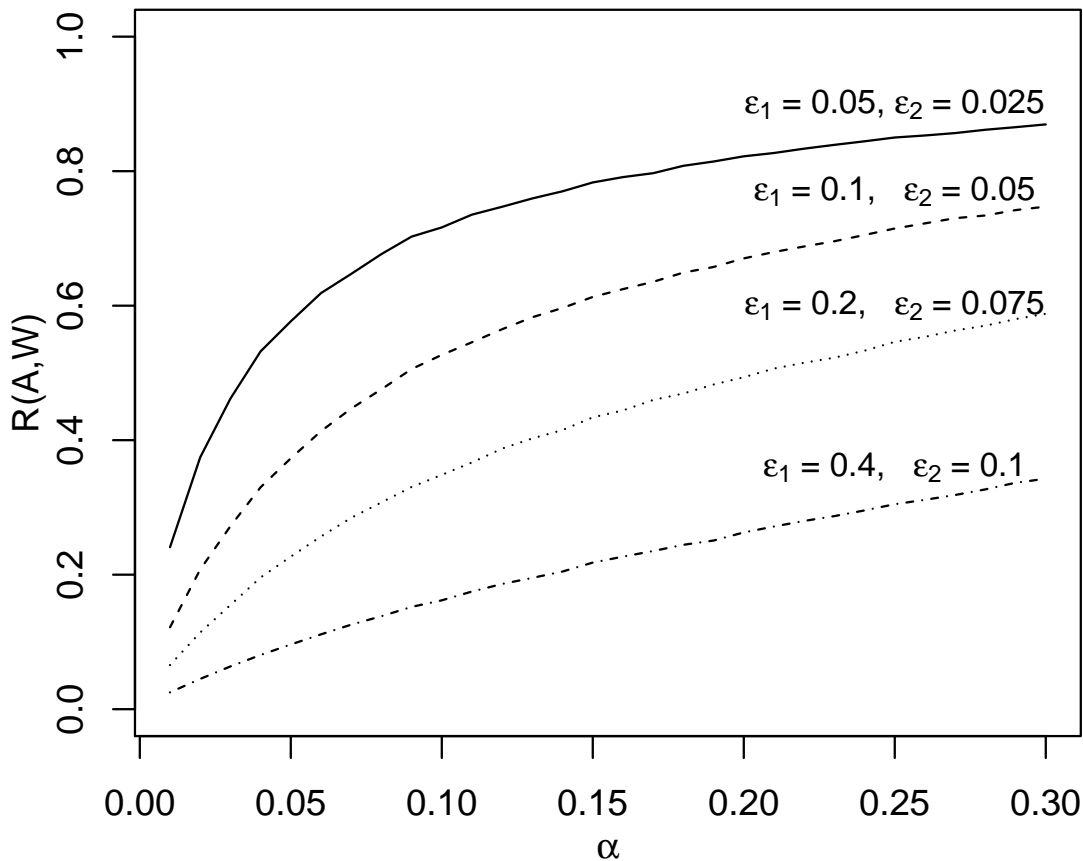}}
	\end{minipage}
	\caption{Ratio of the theoretical variances as a function of $\alpha$ for four choices of $(\epsilon_1, \epsilon_2)$.}
	\label{fig1}	
\end{figure}

For four selected pairs $(\epsilon_1,\epsilon_2)$, Fig.~\ref{fig1} plots $R(\mathbf{A},\mathbf{W})$ as a function of $\alpha$ that is used in creating $\mathbf{A}$. The curves display the averages of ten values given by different $\mathbf{W}$'s. As expected, the efficiency loss caused by inaccuracy in the adjacency matrix is the larger, the more sparse the graph is.

\section{CONCLUSION AND DISCUSSION}
\label{sec:concl}

In this paper, the effect of graph adjacency matrix mismatch has been analyzed and graph error models for different types of graphs, e.g.,  directed/undirected and weighted/unweighted graphs, have been developed. The complexity of the error models varies from simple Erd\"{o}s-R\'{e}nyi type model to one which captures nonconstant edge mislearning probabilities. The latter is of interest because it has been reported in GSP literature that deleting one edge can have immensely larger effect than deleting another edge. Better understanding and formalization of what kind of edges are important or why connecting some pairs of nodes is more harmful than others is therefore crucial. 
The graph error models have been applied for studying graph error effects in graph signal filtering and graph signal ICA applications, where both theoretical arguments and numerical studies for real and synthetic data have been used. The results for different application examples differed in whether missing or extra links were more detrimental, and in the GARMA filter example, it was observed that the higher-order filters were more sensitive to graph errors. These findings suggest that the graph error effects need to be studied case by case, and that competing GSP methods may differ in terms of their robustness to graph errors. Therefore, studying the robustness properties of the GSP methods as well as developing robust GSP methods is of high importance for GSP in general.

\appendices

\section{PROOF OF THEOREM 1}

\subsection{Introduction to Graphons}
Limits of sequences (when the number of nodes tends to infinity) of dense graphs (such graphs in which each node is connected to positive percent of the other nodes) cannot be easily defined using graphs with countable set of nodes since there is no uniform distribution on countably infinite sets. Hence, uncountable cardinality of nodes is needed in order to have a natural graph sequence limit object. In this context graphon $W$ is interpreted as a weighted graph with uncountable set of nodes, $W(x,y)$ giving the edge weight between nodes $x$ and $y$. In~\cite{AvellaMedinaetal2018}, centrality measures of graphons were defined and they were shown to be limit objects of centrality measures of finite size graphs.

Another role of graphons is a kernel in the kernel graph model, which consists of the triple $\mathcal{G}=(N,W,\mu)$, where $N$ is the number of nodes, $W$ is a graphon (or kernel) and $\mu$ is a function/mapping from a probability space to the space $[0,1]^N$. 

A realization of a kernel graph is obtained by generating values $u_1,\dots,u_N$ from $\mu$. Then the $i$th and the $j$th nodes are connected with probability $W(u_i,u_j)$ independently from the other pairs. Thus, the graphon can be seen here as sort of a {\it density function for the graph structure}. Typically $\mu$ is chosen to have uniform density on $[0,1]^N$. 

Some variants of the kernel graph model have been discussed in the literature. For example, $N$ can be derived through a random process or $\mu=\mu_D$ can be taken as a deterministic function that gives sequence $u_i=i/N$ for $i=1,\dots,N$ with probability one. For example, the Erd\"{o}s-R\'{e}nyi model with probability parameter $\epsilon$, which is of interest here, is obtained by choosing the constant graphon $W(x,y) = \epsilon$ for all $0 \leq x, y\leq 1$ and for any choice of $\mu$. 

The SBM with fixed community sizes $N_1,\dots,N_r$ is given by the deterministic $\mu_D$ and $W = W_{\rm SB}$ satisfying $W_{\rm SB} (x,y) = p_{k,l}$ when $\sum_{i=1}^{k-1} N_i/N < x \leq \sum_{i=1}^k N_i/N$ and $\sum_{i=1}^{l-1} N_i/N < y \leq \sum_{i=1}^l N_i/N$ where $\sum_{i=1}^0 N_i = 0$. When $\mu$ has the uniform distribution and $W=W_{\rm SB}$, we have a SBM with random community sizes, the expected sizes being $N_1, \dots, N_r$. 

In the kernel graph model, the expected degree of a random node is 
\[
(N-1) \int_0^1 \int_0^1 W(x,y) d \mu(x) d \mu(y)
\] 
and the range of expected degrees is 
\[
(N-1) \int_0^1 W(x,y) d\mu(y), \ x\in [0,1].
\] 

While the stochastic block model allows for a very heterogeneous degree distribution with the extreme setup $r=N$ and $N_1=\cdots=N_N=1$, such construction is quite cumbersome. Therefore, it is of interest to use continuous graphons which only have few parameters, such as the exponential model~\cite{MorencyLeus2017} where the graphon is of the form 
\[
W(x,y) = e^{-(\beta_1 (x+y) + \beta_0)}
\] 
with $\beta_0, \beta_1 \geq 0$. Parameter $\beta_0$ controls here the sparsity of the graph. The larger the value of $\beta_0$ is, the more sparse the graph is. Moreover, increasing the parameter $\beta_1$ makes the degree distribution more heterogeneous. 

The eigenvalues of large normalized and unweighted graph adjacency matrices can be then approximated in terms of the eigenvalues and eigenfunctions of the kernel~\cite{Lovasz2012}. The eigenvalues $\lambda$ and eigenfunctions of the kernel corresponding to graphon $W$ can be found by solving the equation 
\[
\lambda f(x)=\int_0^1 W(x,y)f(y)dy.
\]

Due to symmetry, any eigenvalue $\lambda$ is real-valued. Examples of solving the eigenvalue/eigenfunction problem in the cases of Erd\"{o}s-R\'{e}nyi, stochastic block and exponential models can be found in~\cite{MorencyLeus2017}, where the behavior of the eigenvalues of finite graphs is also studied in terms of simulations. Thus, graphons have been proved to be useful in the analysis of graph sequences, but they are also convenient for formulating other theoretical results. Here, we are rather intrested to justify the use of Erd\"{o}s-R\'{e}nyi graphon as a basic model for modelling graph errors in graph signals.

\subsection{Proof of Theorem~\ref{theorem1}}

Consider the random variable $c_1W_1(x,y)$ where $x$ and $y$ are fixed and $W_1$ is picked randomly from $\mathcal{W}_\epsilon$. First notice that the distribution is the same for all pairs $(x,y)$, and therefore $\E\{c_1W_1(x,y)\}=\E\{c_1\}\E\{W_1(x,y)\}=c\epsilon$. The result then follows by applying the law of large numbers.

\section{DERIVATION OF EXPECTED VALUE OF THE GRAPH AUTOCORRELATION FOR THE EXAMPLE IN SECTION~III}

Let us first compute here the expectation in the first equation in~\eqref{autocorreqns}. The calculations of the expectations in the other two equations are explained afterwards shortly. We start by substituting \eqref{GSmodel} into the expectation in the first equation in \eqref{autocorreqns}, and opening the square
\begin{align*}
&\E\{\|\mathbf{z}\|^2\}= \E\{\|\left(\theta\mathbf{A}\mathbf{y} \!+\! \mathbf{y}\right)\|^2\} \\
=& \E\left\{\sum_{i=1}^N\left(\theta\sum_{k=1}^N a_{ik}y_k \!+\! y_i-\frac{1}{N}\sum_{j=1}^N \left(\theta\sum_{l=1}^N a_{jl}y_l-y_j\right) \right)^2 \! \right\} \\
=& \E\left\{\sum_{i=1}^N\left(\theta\sum_{k=1}^N a_{ik}y_k+y_i-\frac{1}{N}\sum_{j=1}^N \left(\theta\sum_{l=1}^N a_{jl} y_l-y_j \right)  \right) \right. \\
\times& \left.\left(\theta\sum_{k'=1}^N a_{ik}y_k'+y_i-\frac{1}{N}\sum_{j'=1}^N \left(\theta\sum_{l'=1}^N a_{j'l'}y_l'-y_j'\right) \right)
\right\}. 
\end{align*}
Then after some algebraic manipulations of the latter, we obtain that
\begin{align*}
\E\{\|\mathbf{z}\|^2\}
&= \E\left\{\! \sum_{i=1}^N \! \left( \! \theta^2\sum_{k=1}^N a_{ik}^2y_k^2 \!-\! \frac{2}{N} \! \sum_{j=1}^N \! \left( \! \theta^2 \sum_{k=1}^N a_{ik}a_{jk} y_k^2 \!-\! \theta a_{ij}y_j^2\right) \right.\right. \\
& +y_i^2 \!-\! \frac{2}{N}\sum_{j=1}^N \! \left( \theta a_{ji}y_i^2 \!-\! y_iy_j \right) \!+\! \frac{\theta^2}{N^2}\sum_{j=1,j'=1,l=1}^N \!\! a_{jl}a_{j'l}y_l^2 \\
& \left.\left.\left.+ \frac{2\theta}{N^2}\sum_{j=1,l=1}^N a_{jl}y_l^2+\frac{1}{N^2}\sum_{j=1}^N y_j^2\right)\right)
\right\} \\
=& \sigma_y^2 \left( N^2 \theta^2 \alpha a^2 - 2N \theta^2 \alpha a^2 - 2N^2 \theta^2 \alpha^2 a^2 - 2N \theta \alpha a + N \right. \nonumber \\ -&\left. 2N \theta \alpha a  - 2 + N \theta^2 \alpha a^2 + N^2 \theta^2 \alpha^2 a^2 + 2N \theta \alpha a + 1\right).
\end{align*}

Dropping the negligible terms in the above expression, i.e., the zero-order terms of $N$ and first-order terms of $N$ that also include $a^2$, the first equation in \eqref{autocorreqns} can be approximated as 
\begin{align}
\label{Der2} \nonumber 
\frac{1}{N} \E\{\|\mathbf{z}\|^2\} \approx \sigma_y^2\left((\alpha-\alpha^2) \theta^2a^2N-2\alpha \theta a+1\right).
\end{align}

For deriving the expected values in the other two equations of \eqref{autocorreqns}, there are sums over terms of the type $w_{ij}w_{i'j'}a_{kl}a_{k'l'}y_m^2$. To calculate the expected values of them, we can use the fact that $w_{ij}$ and $a_{kl}$ are independent if $\text{min}(i,j)\neq \text{min}(k,l)$ or $\text{max}(i,j)\neq \text{max}(k,l)$, and then 
\begin{align*}
\mathbb{P}(w_{ij}=w \; {\rm and} \; a_{kl}=a) &= \mathbb{P}(w_{ij}=w) \mathbb{P}(a_{kl}=a) = (\alpha(1-\epsilon_1) + (1 - \alpha) \epsilon_2))\alpha.
\end{align*}
If $\text{min}(i,j)= \text{min}(k,l)$ and $\text{max}(i,j)= \text{max}(k,l)$, we have
\[
\mathbb{P}(w_{ij}=w \; {\rm and} \; a_{kl}=a)=(1-\epsilon_1)\alpha.
\]


\label{sec:ref}

\end{document}